%% file: Review_of_control_techniques_in_aquaculture.tex
\documentclass[a4paper,fleqn]{cas-dc}
\usepackage[numbers]{natbib}
\usepackage[thinlines]{easymat} 
\usepackage{amsmath,amsfonts,amssymb} 
\usepackage{indentfirst}
\usepackage{graphicx}
\usepackage{subeqnarray}
\usepackage{mdwlist}
\usepackage[percent]{overpic}
\usepackage{epstopdf}
\usepackage{rotating}
\usepackage{xcolor}
\usepackage{siunitx}
\usepackage{multirow}
\definecolor{amethyst}{rgb}{0.6, 0.4, 0.8}
\usepackage{geometry}
\usepackage{fleqn.clo}
\usepackage{hyperref}

\usepackage{caption}
\usepackage{subcaption}
\usepackage{pifont}
\definecolor{green2}{RGB}{0,128,0}
\usepackage[utf8]{inputenc}
\usepackage{booktabs,lipsum}
\usepackage[ruled, norelsize, linesnumbered]{algorithm2e}
\usepackage{algpseudocode}
\usepackage{multirow}
\usepackage[switch]{lineno} 
\input{setup-IEEEtrans.tex}

\newcommand{\removelatexerror}{}
\def\tsc#1{\csdef{#1}{\textsc{\lowercase{#1}}\xspace}}
\tsc{WGM}
\tsc{QE}
\tsc{EP}
\tsc{PMS}
\tsc{BEC}
\tsc{DE}

\begin{document}
\let\WriteBookmarks\relax
\def\floatpagepagefraction{1}
\def\textpagefraction{.001}
\shorttitle{Feeding control and water quality monitoring in aquaculture systems}
\shortauthors{F. Aljehani et~al.}


\title [mode = title]{Feeding control and water quality monitoring in aquaculture systems: Opportunities and challenges}

\tnotemark[1]

\tnotetext[1]{This work was supported by the King Abdullah University of Science and Technology (KAUST) Base Research Fund (BAS/1/1627-01-01).}


\author[1]{Fahad Aljehani}[]
            
\ead{fahad.aljehani@kaust.edu.sa}

\author[1]{Ibrahima N'Doye}[]

\ead{ibrahima.ndoye@kaust.edu.sa}

\author[1,2]{Taous~Meriem Laleg-Kirati}[]

\ead{taousmeriem.laleg@kaust.edu.sa}

\address[1]{Computer, Electrical and Mathematical Sciences and Engineering Division (CEMSE), King Abdullah University of Science and Technology (KAUST), Thuwal 23955-6900, Saudi Arabia}
\address[2]{National Institute for Research in Digital Science and Technology, Paris-Saclay, France.}

\cortext[cor1]{Corresponding author: T. M. Laleg-Kirati}

\begin{abstract}
Aquaculture systems can benefit from the recent development of advanced control strategies to reduce operating costs and fish loss and increase growth production efficiency, resulting in fish welfare and health. Monitoring the water quality and controlling feeding are fundamental elements of balancing fish productivity and shaping the fish growth process. Currently, most fish-feeding processes are conducted manually in different phases and rely on time-consuming and challenging artificial discrimination. The feeding control approach influences fish growth and breeding through the feed conversion rate; hence, controlling these feeding parameters is crucial for enhancing fish welfare and minimizing general fishery costs. The high concentration of environmental factors, such as a high ammonia concentration and pH, affect the water quality and fish survival. Therefore, there is a critical need to develop control strategies to determine optimal, efficient, and reliable feeding processes and monitor water quality. This paper reviews the main control design techniques for fish growth in aquaculture systems, namely algorithms that optimize the feeding and water quality of a dynamic fish growth process. Specifically, we review model-based control approaches and model-free reinforcement learning strategies to optimize the growth and survival of the fish or track a desired reference live-weight growth trajectory. The model-free framework uses an approximate fish growth dynamic model and does not satisfy constraints. We discuss how model-based approaches can support a reinforcement learning framework to efficiently handle constraint satisfaction and find better trajectories and policies from value-based reinforcement learning.
\end{abstract}


\maketitle
\section{Introduction}
\begin{figure*}[!t]
\centering
   \begin{overpic}[scale=0.28]{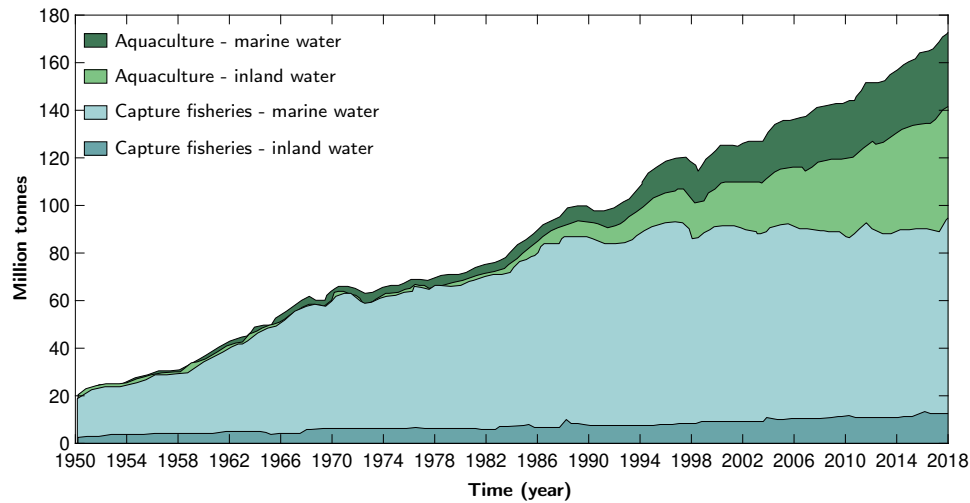}
   \end{overpic}\vspace{-0.1cm}
   \put(-352,170){\scriptsize Aquaculture - marine water}
  \put(-352,157){\scriptsize Aquaculture - inland water}
  \put(-352,144){\scriptsize Capture fisheries - marine water}
  \put(-352,130){\scriptsize Capture fisheries - inland water}
  \put(-390,75){\scriptsize \rotatebox{90}{\textbf{Million tonnes}}}
  \put(-220,2){\scriptsize {\textbf{Time (year)}}}
   \caption{Marine and aquaculture worldwide production.}\label{roadmap} 
 \end{figure*}

Fisheries play a vital role in the global food supply and are becoming critical components in countries' economies \cite{CaL:15}. In its latest world fisheries and aquaculture report, the Food and Agriculture Organization stated that the fishery product deficit will reach $29$ million tons in 2030 \cite{FAO:15, FAO:18}. Wild capture cannot cover this amount; therefore, the fishery market is turning toward aquaculture production, which has become a key provider of seafood and a crucial element for sustainable food production \cite{FAO:15, FAO:19}, as illustrated in Figure~\ref{roadmap}. Fish farming in the sea has undergone several developments, including the design of sea cages with useful properties (e.g., submersible or floating), the development of reproduction and growth technologies for some species, and the design of appropriate physical arrangements for cultured organisms \cite{FAO:15, FAO:19}.

Direct human observations can build a relationship between farmers and fish, influencing fish welfare, health, and productivity. However, due to the fish population and external factors, creating such a regime based on observations alone might be inadequate for building relationships. Therefore, technological tools are needed to monitor the fish populations remotely so that the data can be used to regulate routine operations and optimize fish growth and survival \cite{Fo:18}.

The feeding process and water quality monitoring are typically performed manually, on-site, or in laboratories after the data are collected \cite{SHOLS:19}. This monitoring process delays the detection of abnormalities and relevant control actions and is arduous for managing the cleaning costs and complexity and stabilizing the water quality \cite{Fo:18}. Recent technological advances have helped improve aquaculture monitoring, enabling continuous remote monitoring of fish health and well-being in various aquaculture environments \cite{DaE:09, TBJAPC:09, Fo:18}. Additionally, deploying technology helps minimize losses while increasing productivity. In this context, wireless sensor network tools and advanced control strategies have been recently investigated. They allow more efficient data collection and reduce the operating costs for monitoring water quality \cite{RRM:16} and evaluating daily fish feeding \cite{VSLG:13, DNBBF:16}. Moreover, they serve as an appropriate way to transmit data and information to the platforms to initiate specific actions \cite{MKBWS:06, TLMV:16, Fo:18}. Even though technology has helped improve the monitoring process, it still lacks full autonomy for decision-making. From this perspective, the involvement of control engineering in aquaculture production is necessary to improve decision-making and monitoring of the overall health of fish.

In addition, computer vision has been used to develop key algorithms and software solutions to analyze collected data and provide useful metrics to improve decision-making \cite{JBFHJVS:11, SoF:15, ZHW:13, CZYX:17, AnC:20}. Computer vision algorithms allow, for example, fish position and velocity detection, measuring size and weight, and monitoring fish density, all important factors in decision-making \cite{SoF:15}. These metrics can indirectly relate to fish behavior because the fish trajectory can indicate behavior \cite{ANJMBL:22}. Some research studies have proposed analyzing fish behavior based on computer vision-based metrics. However, current data-driven control studies have not been connected to any decision-making process and are still investigating how the feedback control loop can be closed and the direct relation between fish behavior and the feeding process, for example, \cite{MSO:03, AnC:20}.

Fish aquaculture takes care of the life cycle of the fish from the egg to the adult stage in indoor recirculating cages, outdoor ponds, or sea cages \cite{Fo:18}.
With the increasing scale of fish production, optimal management becomes much more critical to satisfying economic efficiency standards. The fish growth rate relies on external factors, including water quality and food, and internal factors, including the nutritional and physiological aspects of the fish. Hence, fish overfeeding increases wasted food costs, affecting water quality and consequently increasing maintenance and cleaning operations. Building an ideal fish culture, feeding, and most other environmental operations (e.g., temperature, oxygen, light intensity, and water flow) could be optimized for the entire cage population, where responses corresponding to specific fish behaviors could indicate particular activity levels, such as hunger or energy. At this stage, no closed-loop examples of fish farming systems consider various observational components and lead to decision-making \cite{Fo:18}. Therefore, the aquaculture industry and research aspire to monitor and control the effects of various external challenges that affect the ability to maintain an appropriate productive level of the fishery and simultaneously respect the diverse biological and environmental characteristics. In addition to the previously mentioned challenges, aquaculture monitoring faces challenges related to sensing and communication \cite{Fo:18}. Developing advanced control strategies to address these challenges can support reducing management costs and fish loss, enhancing production efficiency, and improving fish quality (health and welfare).
 
Control systems have been applied to diverse branches of engineering and science and drive a long history of mathematical rigor. Generations of scientists and engineers have been developing control methods, tools, and algorithms to solve several practical issues of importance with tremendous influence on society \cite{BaG:11}. Control concepts have been essential in designing and developing industrial process plants, high-performance airplanes, manufacturing enterprises, fuel-efficient automobiles, communication networks, smartphones, planetary rovers, and other applications across various industry sectors. The applicability of control concepts is gaining the trust of researchers and engineers outside the control field due to their reliable, efficient, and cost-effective operations in practical applications.

Recently, advanced control techniques for fish growth tracking have established significant possibility and resilience to fulfill the fish farmers' markets depending on the fish species, feed cost, culture duration, and selling price \citep{CNMBL:21}. However, increasing the complexity of aquaculture systems can lead to model uncertainty, especially in floating cages. The presence of uncertainties imposes challenges on model-based controllers. Specifically, the well-known representative fish growth model based on energy-balance models is sensitive to variations in external environmental factors. Therefore, modeling fish growth trajectories with the presence of uncertainties is a crucial task. 

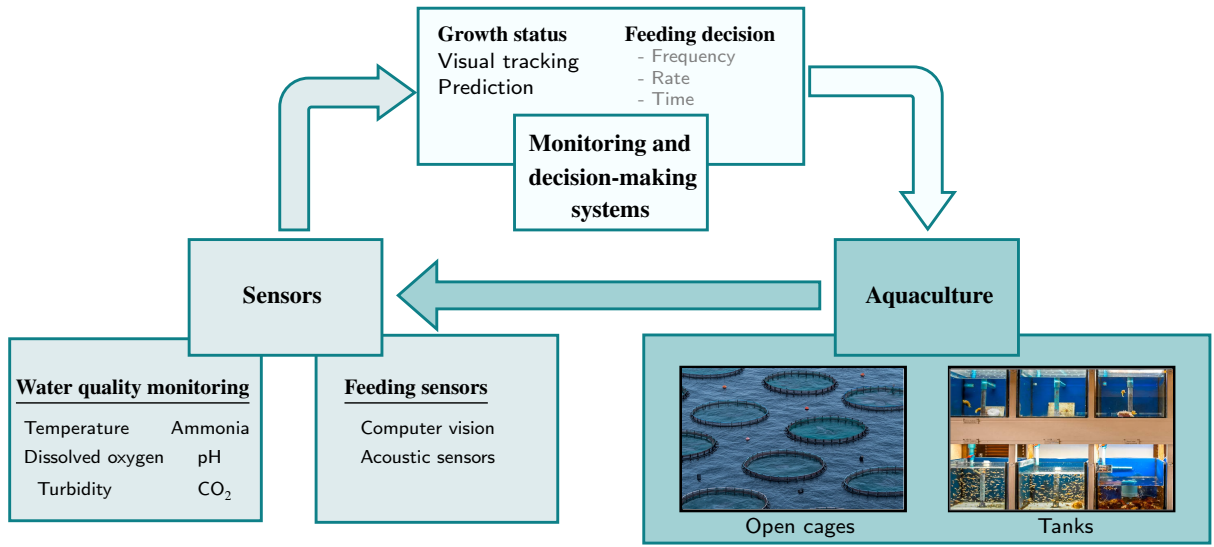
\begin{figure*}[!t]
\centering
   \begin{overpic}[scale=0.5]{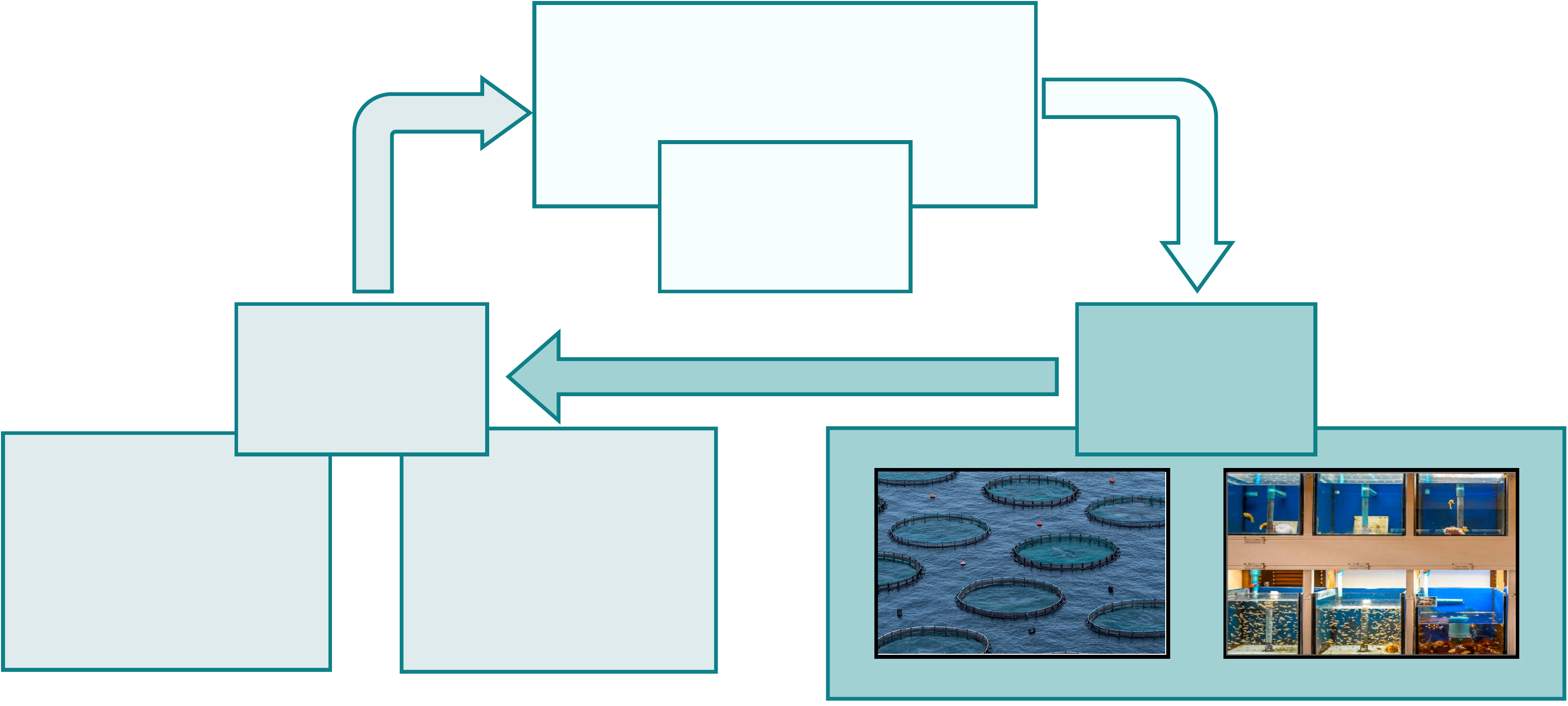}
   \end{overpic}
   \put(-130,92){\bf Aquaculture}
   \put(-175,6){\footnotesize Open cages}
   \put(-65.5,6){\footnotesize Tanks}
   \put(-363,92){\bf{Sensors}}
   \put(-448,58){\footnotesize \underline{\bf Water quality monitoring}}
   \put(-445,43){\scriptsize Temperature}
   \put(-390,43){\scriptsize Ammonia}
   \put(-445,32){\scriptsize Dissolved oxygen}
   \put(-380,32){\scriptsize pH}
   \put(-440,20){\scriptsize Turbidity}
   \put(-380,20){\scriptsize $\mbox{CO}_2$}
   \put(-325,58){\footnotesize \underline{\bf Feeding sensors}}
   \put(-319,43){\scriptsize Computer vision}
   \put(-319,32){\scriptsize Acoustic sensors}
   \put(-256,149){\bf Monitoring and}
    \put(-256,136){\bf decision-making}
      \put(-240,124){\bf systems}
   \put(-290,190){\footnotesize \bf Growth status}
   \put(-290,180){\footnotesize Visual tracking}
   \put(-290,170){\footnotesize Prediction}
   \put(-220,190){\footnotesize \bf Feeding decision}
   \put(-215,182){\scriptsize \textcolor{gray}{- Frequency}}
   \put(-215,174){\scriptsize \textcolor{gray}{- Rate}}
   \put(-215,166){\scriptsize \textcolor{gray}{- Time}}
   \caption{Smart aquaculture sensors and monitoring and decision-making systems.}\label{sensing-monitoring} 
 \end{figure*}

The learning-based method for fish growth, such as reinforcement learning (RL), can alleviate the model uncertainty effects. The Q-learning algorithm for fish trajectory tracking is a model-free RL algorithm, which was introduced in \cite{CNMBL:22}. Specifically, the Q-learning algorithm controller aims to track the fish growth trajectory while managing model uncertainties and environmental factors, such as dissolved oxygen and water temperature. This paper aims to review reliable decision support and automation systems based on continuous water quality monitoring and feeding control for fish growth. It also aims to improve accuracy and precision in farming operations to remotely observe fish-feeding habits and growth status, as illustrated in Figure~\ref{sensing-monitoring}. This paper can help continuously predict and control the effect of environmental factors and feed quality factors to improve the ability of farmers to document biological processes in farm-based fish production.

This article is organized as follows. Section~\ref{feeding-water} explores the various components, such as water quality and feeding, that use feedback loops in aquaculture systems. Next, Section~\ref{background} introduces general and bioenergetic fish growth and the challenges of building these fish growth models in aquaculture systems. Then, Section~\ref{Model-based} addresses more advanced model-based control systems, such as model predictive control (MPC). Section~\ref{Model-free} details the advantage of deriving a model-free controller based on an RL approach to achieve better control performance. In addition, Section~\ref{Integrated-model} describes promising algorithms for simultaneously designing model-based and learning controllers and highlights opportunities in the aquaculture field. Finally, Section~\ref{conclusion} provides concluding remarks and future directions for research.

\section{Effects of feeding and water quality on growth and survival}\label{feeding-water}
This section introduces the definition of feeding and water quality in aquaculture and their effects on fish growth. A variety of threshold ranges exist depending on the kind of fish. However, understanding their effects on growth plays an essential role from a control perspective. 

\subsection{Feeding}
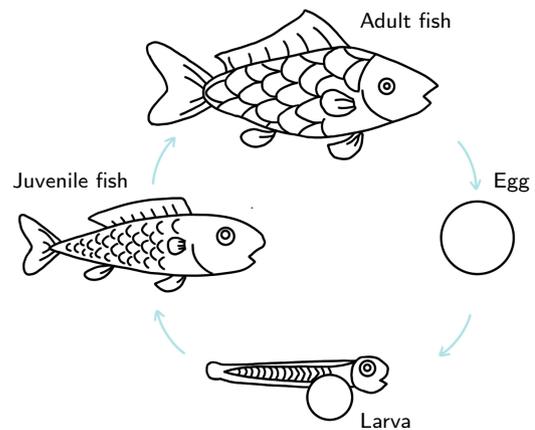
\begin{figure}[!h]
\vspace{0.25cm}
\centering
   \begin{overpic}[scale=0.25]{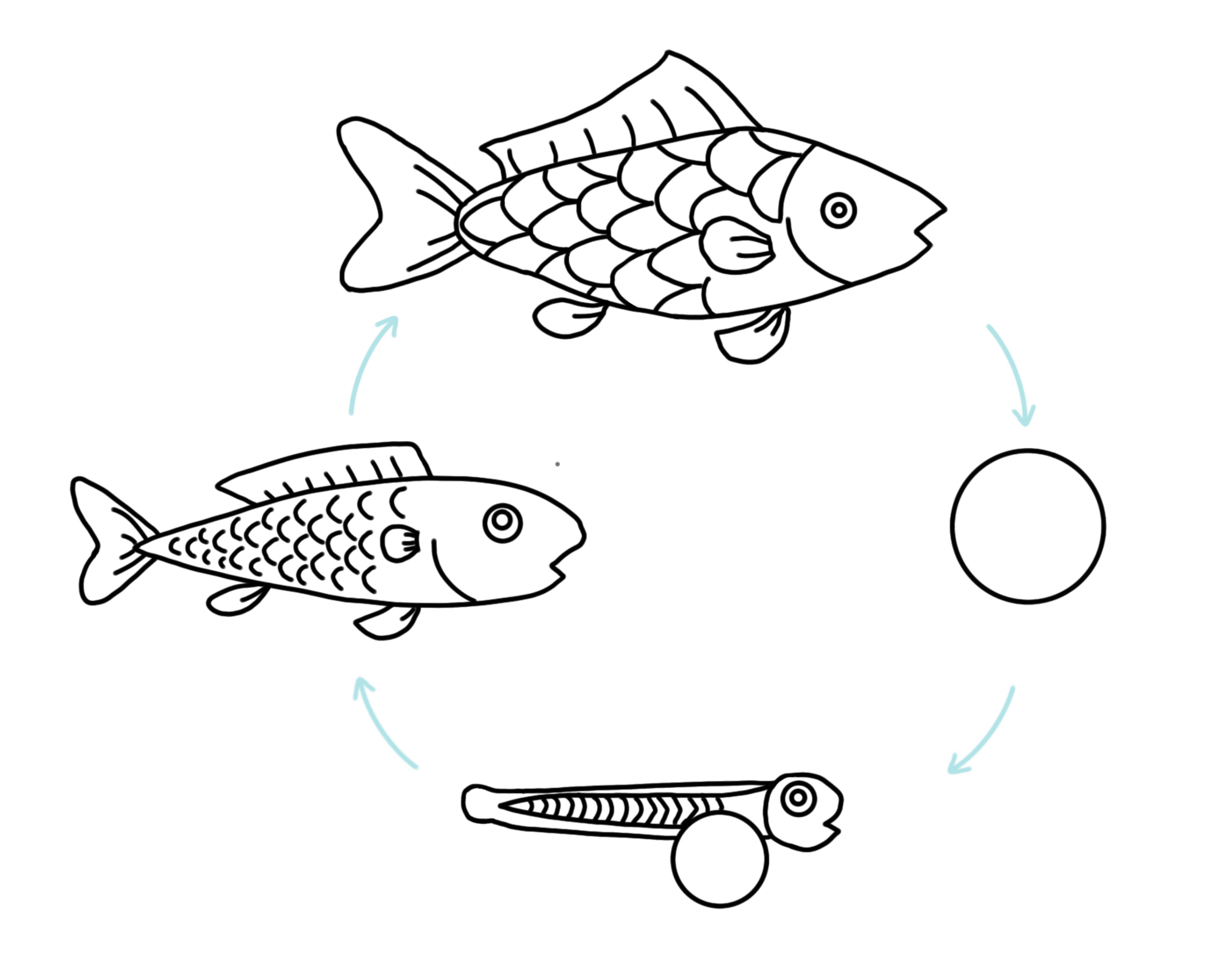}
   \end{overpic}\vspace{0.3cm}
   \put(-80,160){\footnotesize Adult fish}
   \put(-210,100){\footnotesize Juvenile fish}
   \put(-30,100){\footnotesize Egg}
   \put(-80,10){\footnotesize Larva}
  \vspace{-0.5cm}
   \caption{Fish life cycle from an egg stage to adult. }\label{FishCycle}
 \end{figure}

Feeding refers to the food provided to fish to maintain a healthy and sustainable life. Food contains proteins, carbohydrates, lipids, vitamins, and minerals. The percentage of these mixtures varies from one species to another, depending on age. Finding a suitable range of these mixtures helps optimize fish growth during the grow-out cycle. Another concern arises regarding the best time and how to feed fish. This concern is essential because, even when providing the fish with the perfect food mixture, they could eventually be overfed or underfed. Figure~\ref{FishCycle} presents the fish life cycle from being an egg to an adult. In each life cycle stage, the concepts of feeding frequency, rate, and time vary, and investigating these three concepts can help avoid overfeeding and underfeeding.

Next, we briefly describe the three feeding concepts and emphasize their importance for fish growth and survival. 
 
\noindent{\textbf{Feeding frequency:}} The concept of feeding frequency refers to how many times the fish feed per day. Depending on size (or stomach size, in particular), the food held in a fish's stomach varies from the juvenile to the adult stages \cite{DavisRonald2022}. Increasing the feeding frequency, to a limited extent, enables the fish to consume the food and grow better \cite{DwyerJoseph2002, WangLing2007}.

\noindent{\textbf{Feeding time:}} Feeding time refers to what time to feed the fish so they grow faster and healthier. Once the feeding frequency is determined, the next objective is to determine the best time to feed the fish. Various studies have mentioned that feeding time could affect fish growth performance \cite{NoeskeSpieler1984, BoujardJourdan1996}. In some cages, feeding at night resulted in higher and better fish growth than morning feeding given the same amount of food \cite{KerdchuenLegendre1991}.

\noindent{\textbf{Feeding rate:}} The feeding rate is the amount of food calculated by the percentage of fish body weight per day. The percentage varies from one species to another, but the most common rates range from 1\% to 10\% of the fish body weight \cite{Zahrani2013, Khandan2019}. Besides the effect on fish growth, increasing the feeding rate can also affect the water quality \cite{Singh2003}.

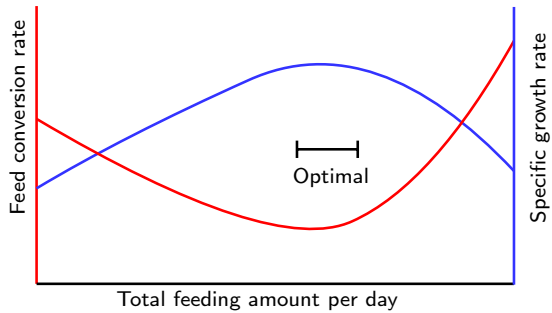
\begin{figure}[!h]
\centering \vspace{0.1cm}
   \begin{overpic}[scale=0.45]{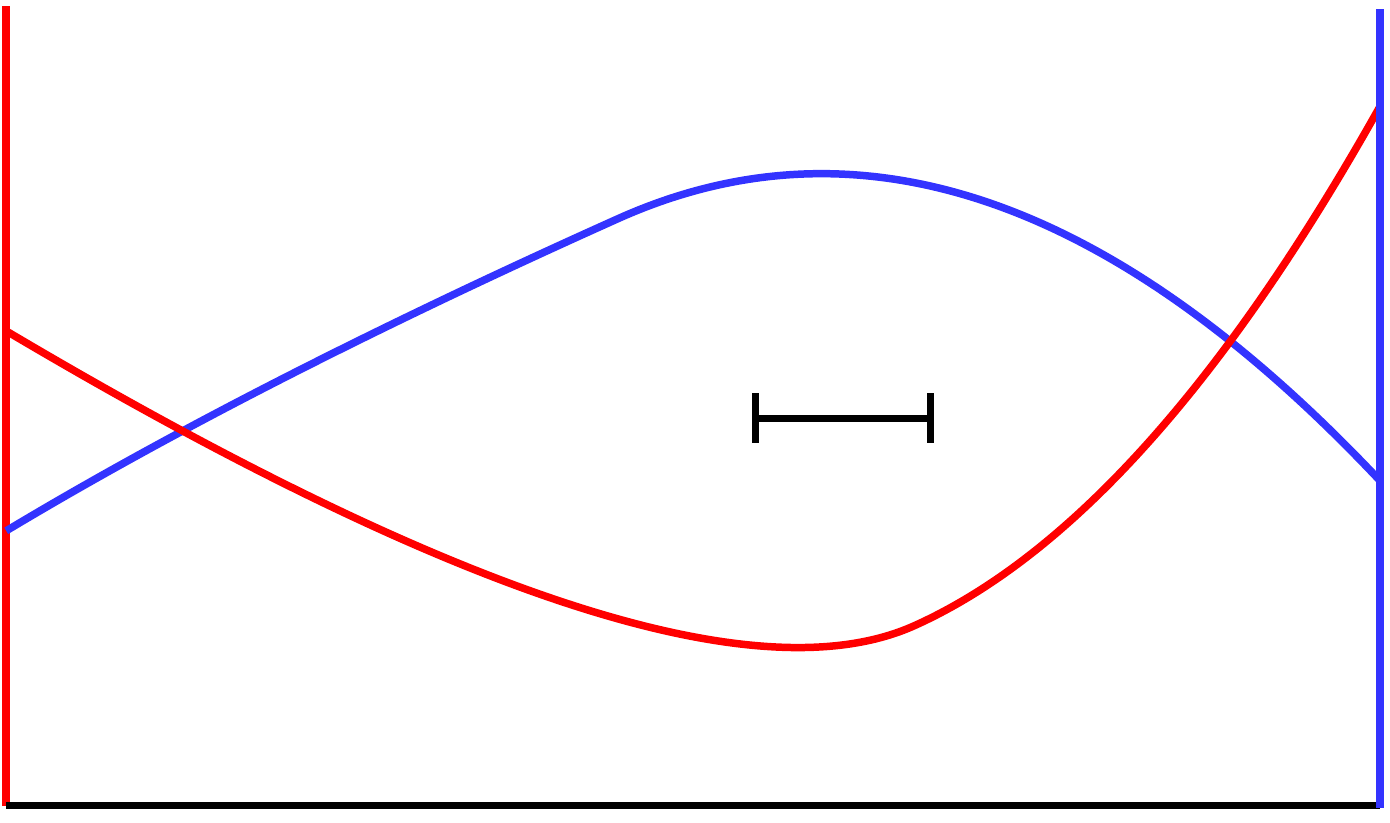}
   \end{overpic}
   \put(-190,28){\footnotesize \rotatebox{90}{Feed conversion rate}}
   \put(5,25){\footnotesize \rotatebox{90}{Specific growth rate}}
   \put(-84,40){\footnotesize Optimal}
   \put(-150,-6){\footnotesize Total feeding amount per day}
   \vspace{0.1cm}
   \caption{Effects of the feed conversion rate and specific growth rate metrics on the total feeding amount.}\label{metrics} 
 \end{figure}

Aquaculturists usually refer to different measuring metrics to test the effectiveness of applying the above three concepts. The feed conversion rate is one metric to determine the feed needed to obtain the desired weight of the fish. Another metric is the specific growth rate, which measures the increased percentage of fish weight over time. Figure~\ref{metrics} illustrates the relationship between these rates on the total daily feeding amount.
 
\subsection{Water quality}
Water quality in culture systems is any characteristic of water that can affect fish growth, production, and survival. Many water quality variables can affect fish health and growth. However, several variables play a vital role; for example, water temperature and salinity become critical variables when assessing a suitable environment for breeding. Dissolved oxygen, ammonia, and carbon dioxide ($\mbox{CO}_2$) play a significant role during growth \cite{Boyd1998, Boyd1985, BOYD2017}. The main objective is to create a sustainable environment for fish to be healthier and produce more efficiently. To this end, this section introduces some water quality variables that can be controlled directly or indirectly from a control theory perspective and details their effects on the growth status. 

\begin{figure}[!h]
\centering
   \begin{overpic}[scale=0.5]{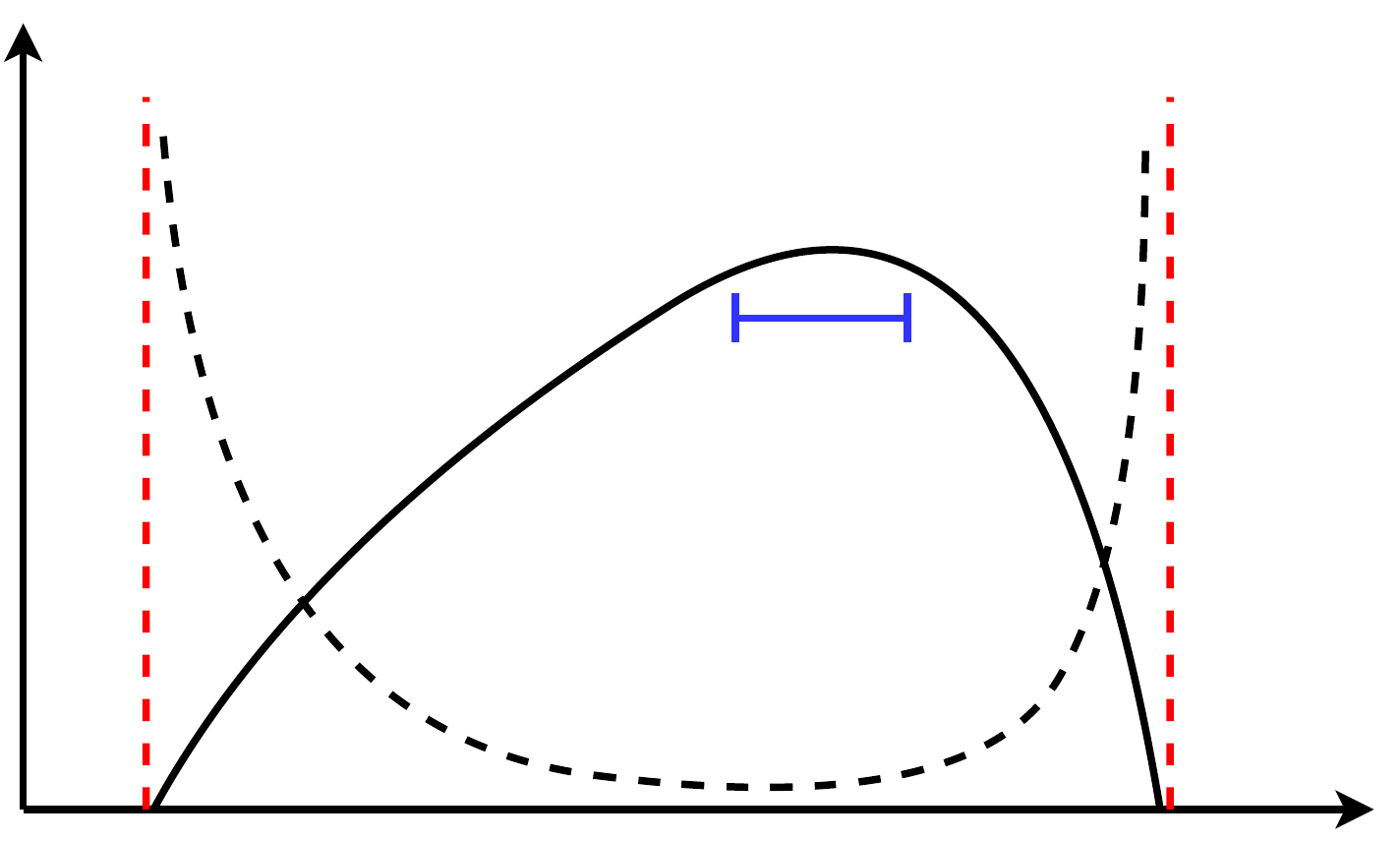}
   \end{overpic}
   \put(-215,45){\scriptsize \rotatebox{90}{\footnotesize Fish growth}}
   \put(-195,25){\footnotesize \rotatebox{90}{Lower lethal boundary}}
   \put(-30,25){\footnotesize \rotatebox{90}{Upper lethal boundary}}
   \put(-125,-6){\footnotesize Temperature}
   \put(-99,65){\footnotesize Optimal}
   \put(-110,15){\scriptsize Mortality}
   \vspace{0.1cm}
   \caption{Effects of temperature on fish growth.}\label{Temperature} 
 \end{figure}
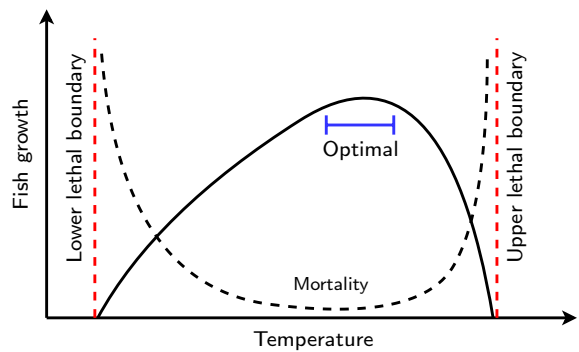

\noindent{\textbf{Temperature:}} Fishes are ectothermic (poikilothermic or cold-blooded) organisms that cannot control their body temperature but equilibrate it with the surrounding water \cite{ProsserHazel1974, MoyleCech2000}. Due to biochemical rates, temperature directly affects the physiology of the fish, including the fish growth rate, activity, reproduction, and oxygen demand \cite{FryHart1948, Ivleva1980, John1975, Schmidt1990}. Regardless of the culture system type, fish species fall into four categories: tropical, warm water, cool water, and cold water. 
The effect of temperature on fish growth is illustrated in Figure~\ref{Temperature}. 

\begin{figure}[!h]
\centering\vspace{0.2cm}
   \begin{overpic}[scale=0.6]{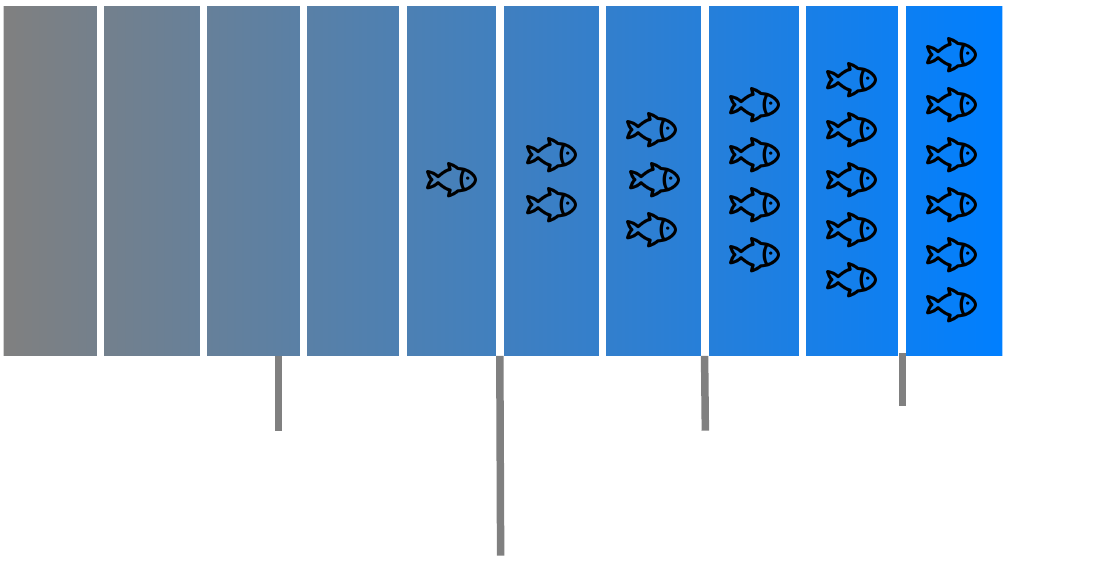}
   \end{overpic}\vspace{0.1cm}
   \put(-193,100){\scriptsize 0}
   \put(-175,100){\scriptsize 1}
   \put(-157,100){\scriptsize 2}
   \put(-139,100){\scriptsize 3}
   \put(-122,100){\scriptsize 4}
   \put(-105,100){\scriptsize 5}
   \put(-87,100){\scriptsize 6}
   \put(-70,100){\scriptsize 7}
   \put(-53,100){\scriptsize 8}
   \put(-36,100){\scriptsize 9}
   \put(-22,100){\scriptsize 10}
   \put(-192,20){\scriptsize Lethal for fish}
   \put(-192,10){\scriptsize populations}
   \put(-140,20){\scriptsize Stressful}
   \put(-140,10){\scriptsize conditions}
   \put(-103,20){\scriptsize Supports}
   \put(-103,10){\scriptsize growth and}
   \put(-103,0){\scriptsize activity}
   \put(-55,20){\scriptsize Supports}
   \put(-55,10){\scriptsize a healthy}
   \put(-55,0){\scriptsize fish population}
   \caption{Effects of dissolved oxygen in parts per million (ppm) on fish growth.}\label{Oxy} 
 \end{figure}
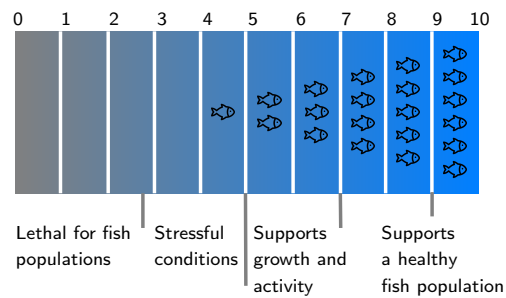

\noindent{\textbf{Dissolved oxygen ($DO$):}} Disolved oxygen is a measurement of the amount of oxygen dissolved in water. A suitable dissolved oxygen level in water can enhance fish growth \cite{BRETT1979}. More importantly, some studies have varied the dissolved oxygen level and noted that less dissolved oxygen than a suitable range affects fish appetite, metabolism, feed efficiency, and feed intake \cite{CuiWootton1988, Jobling1993, CUENCO1985191}. The effect of the dissolved oxygen on fish growth is presented in Figure~\ref{Oxy}.

\begin{figure}[!h]
 \vspace{0.5cm}
\centering
   \begin{overpic}[scale=0.6]{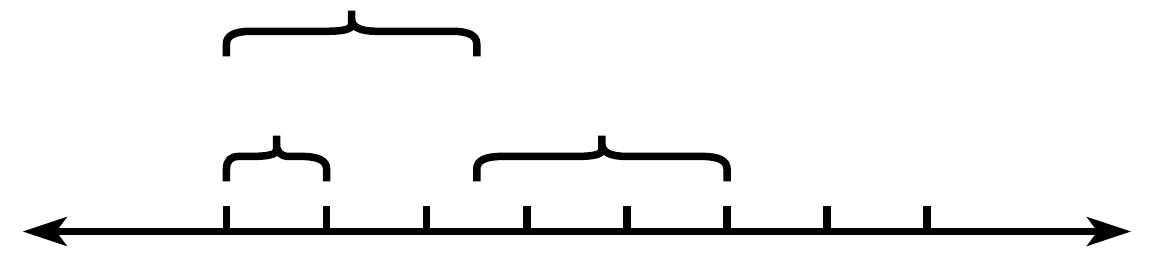}
   \end{overpic}
   \put(-163,-2){\scriptsize 4}
   \put(-145,-2){\scriptsize 5}
   \put(-128,-2){\scriptsize 6}
   \put(-110,-2){\scriptsize 7}
   \put(-93,-2){\scriptsize 8}
   \put(-76,-2){\scriptsize 9}
   \put(-60,-2){\scriptsize 10}
   \put(-43,-2){\scriptsize 11}
   \put(-180,25){\footnotesize Less production}
   \put(-120,25){\footnotesize Suitable range}
   \put(-160,48){\footnotesize Slow growth}
   \put(-105,-10){\footnotesize pH}
   \vspace{0.1cm}
   \caption{Effects of pH on fish growth.}\label{pH-Range} 
 \end{figure}
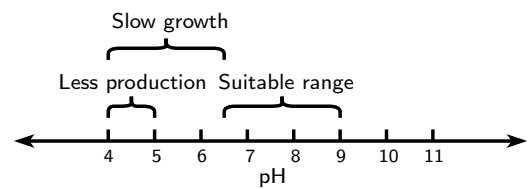

\noindent{\textbf{Potential of hydrogen (pH):}} This is a measurement of whether the water is acidic or basic. Exposure to high ($\text{pH}>> 9$) or low ($\text{pH} <<4.5$) levels causes fish stress and death at the extremes. The suitable range of pH level for most fish for fish growth and production is between 6.5 and 9. The range between 4 and 6.5 can result in less fish production and slow growth. Figure~\ref{pH-Range} summarizes the pH levels and their results \cite{boyd1990}. 
 
\noindent{\textbf{Other factors:}} Other factors, such as ammonia, $\mbox{CO}_2$, and turbidity can affect water quality. Exceeding the suitable range of ammonia concentration and excessive $\mbox{CO}_2$ result in poor feed conversion that consequently reduces fish growth performance and increases mortality \cite{Shin2016}, stimulating the stress response \cite{Vasco2019}. Furthermore, increasing turbidity affects the temperature and dissolved oxygen, reducing fish growth \cite{Reed1983}. 

\section{Fish growth models in aquaculture systems}\label{background}
This section explains fish growth models, how they are derived, and the challenges of building a representative bioenergetic fish growth model, which allows for designing model-based controllers for the growth trajectory problem. We start with a brief introduction to general growth functions, providing an appreciation of the state development of a fish and its environment (temperature, food, and water quality), underlying fish growth performance.

\subsection{Fish growth models}
Growth models describe how fish respond to the surrounding environment and the rich interactions between fish weight, feeding ratio, and temperature that coevolve over time in marine recirculation and open-cage aquaculture systems. Most growth models are derived from biological and empirical characteristics. These characteristics are described in mass-balance or energy-balance models, where the growth rate is expressed as metabolism. Metabolism is a term that describes all biochemical reactions in cells to convert food into energy. 

\begin{figure}[!t]
\vspace{0.1cm}
 \centering
 \subfloat{\includegraphics[width=0.4\textwidth]{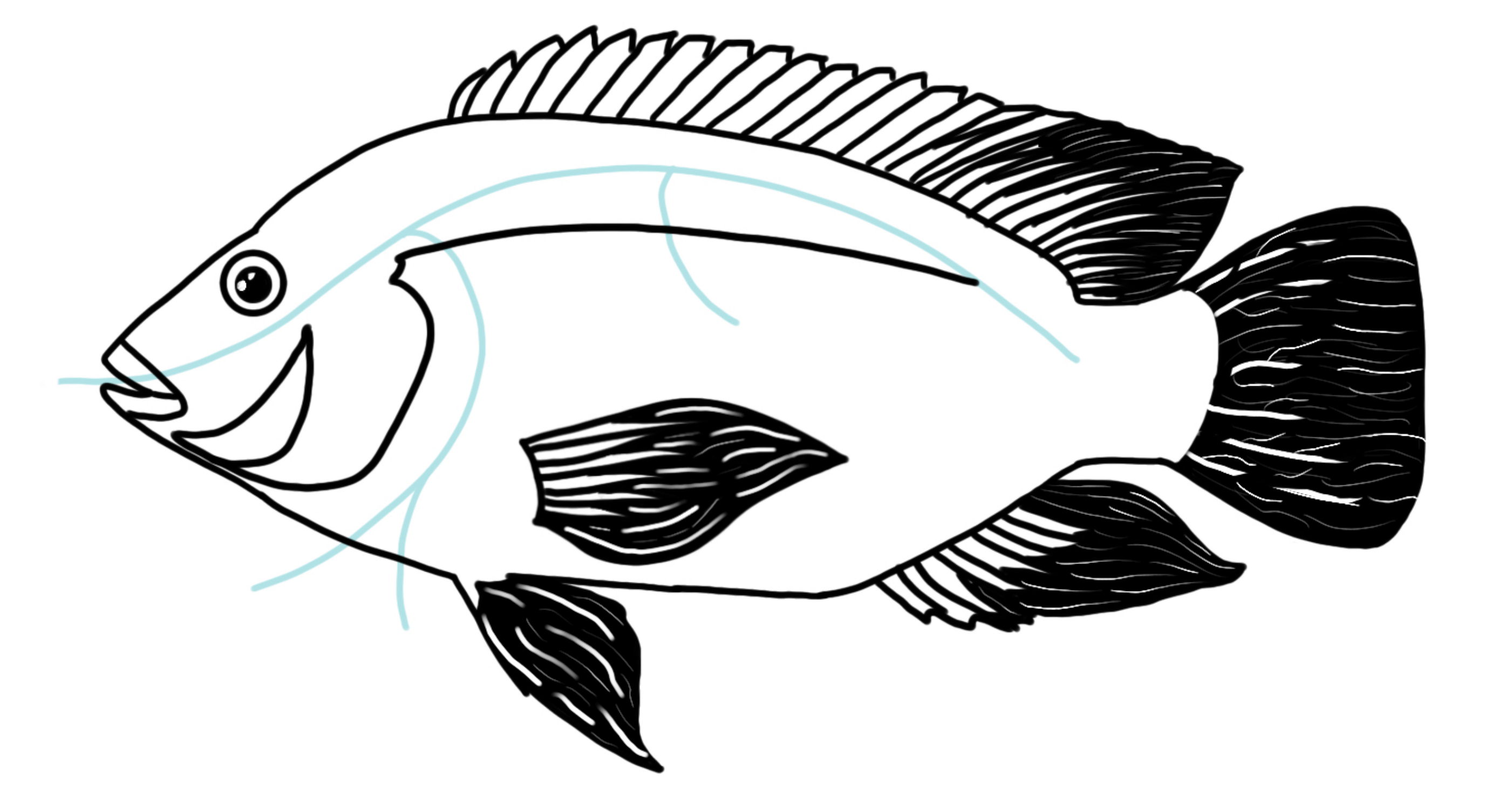}}
 \put(-200,-20){\begin{overpic}[scale=0.15]{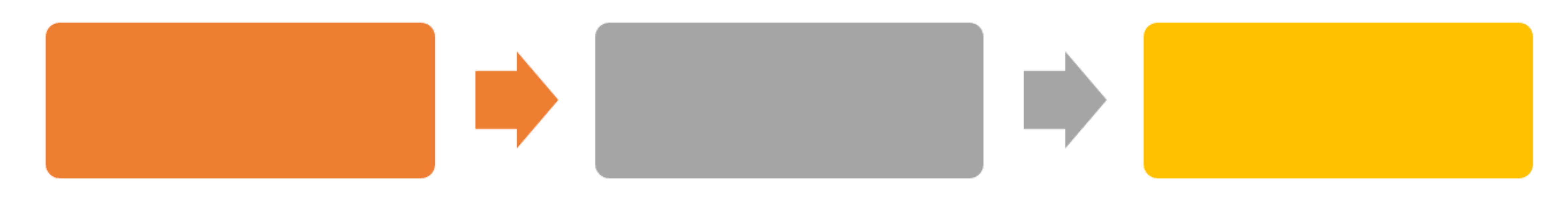}
   \put(5,5){\scriptsize Energy intake}
  \put(45,7){\scriptsize Energy}
  \put(44,3){\scriptsize allocation}
  \put(79,5){\scriptsize Growth}
   \end{overpic}}
  \put(-220,45){\scriptsize \textcolor{gray}{Food intake}}
    \put(-200,86){\scriptsize a)}
  \put(-220,56){\scriptsize Oxygen}
  \put(-120,55){\scriptsize Foraging}
  \put(-65,50){\scriptsize \textcolor{gray}{Growth}}
  \put(-176,20){\scriptsize \textcolor{gray}{SMR}}
  \put(-150,16){\scriptsize SDA}
  \put(-30,80){\scriptsize Locomotion}
  \put(-81,10){\scriptsize Fecal waste}
\vspace{0.5cm}
 \subfloat{\includegraphics[width=0.27\textwidth]{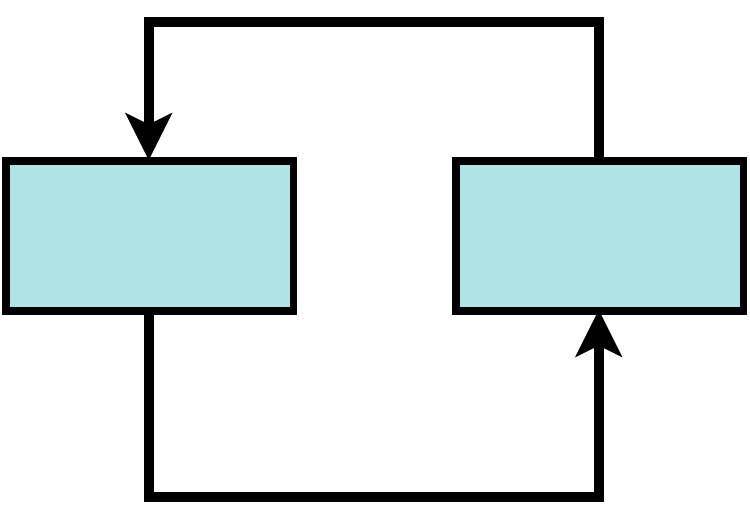}}
      \put(-160,86){\scriptsize b)}
  \put(-128,50){\footnotesize Amino acids}
  \put(-40,50){\footnotesize Proteins}
  \put(-85,76){\footnotesize \textcolor{gray}{Catabolism}}
  \put(-85,10){\footnotesize \textcolor{gray}{Anabolism}}
  \put(-100,95){\footnotesize Energy release adenosine triphosphate}
  \put(-100,-3){\footnotesize Energy consume adenosine diphosphate}
  \vspace{0.3cm}
 \caption{Process of anabolism and catabolism in fish metabolism.}
\label{AnabolismCatabolism}
\end{figure}
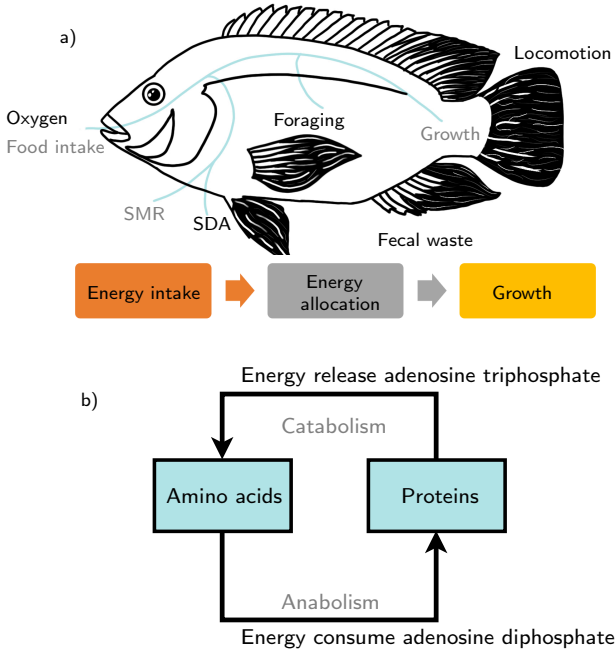

\subsubsection{General fish growth models}
Fish weight, feed ratio, and water temperature are the dominant predictors of fish growth function and preserve most of the essentials of various models. Dynamic growth models are essential for analyzing, predicting, and understanding growth behavior and survival for rational aquaculture management. Growth models for aquaculture often consist of state and environmental vectors. The state factors usually refer to an individual fish's body mass, such as energy and protein content. Environmental vectors include feeding and water quality factors, such as temperature and oxygen. Many growth models \cite{Koo:10, Bon:20, DFB:10, Seg:16} have been proposed to predict the weight of the fish, given the environmental vectors.

In general, the dynamic growth rate system is described as follows \cite{Seg:16}: 
\begin{equation}\label{eq1}
\frac{d\xi}{dt}=\mathbb{G}\Big\{f, T, \xi\Big\},
\end{equation}
where $t$ represents time (normally in days), $\xi$ denotes the body mass of a single fish, $f$ is the feed ration, and $\mathbb{G}$ indicates the growth rate of $\xi$. In addition to $f$ and $T$, the growth function $\mathbb{G}$ may also depend on the dissolved oxygen \cite{Job:94,BGN:00}, ammonia \cite{WGKP:91}, fish density \cite{SRGLBMCFB:09,YiA:10}, $\mbox{CO}_2$ \cite{BSMNL:13}, salinity \cite{KlC:96}, feed composition \cite{LKS:03}, and other factors of interest. 

Equation~\eqref{eq1} is often called a “single-term model” and represents a simplification of the following “two-term models” in which the growth rate is expressed as the difference between anabolism $\mathbb{A}$ and catabolism $\mathbb{C}$ \cite{Seg:16}: 
\begin{equation}\label{eq2}
\mathbb{G}=\mathbb{A}\Big\{f, T, \xi\Big\}-\mathbb{C}\Big\{T, \xi\Big\}.
\end{equation}
The process of metabolic reactions consists of two stages: catabolism and anabolism \cite{JudgeDodd2020}. The catabolism stage is the process of breaking down complex macromolecules into simple molecules. Examples of such processes are breaking down proteins into amino acids, lipids into fatty acids, and carbohydrates into sugars. Usually, this process releases energy in terms of adenosine triphosphate due to the breakdown of the phosphodiester bonds. In contrast, the anabolism stage builds up simple molecules into complex macromolecules, which can be considered the reverse of the catabolism stage. Unlike catabolism, the anabolism stage consumes energy in the form of adenosine diphosphate. Figure~\ref{AnabolismCatabolism} illustrates the correlation between anabolism and catabolism in fish.

\subsubsection{Bioenergetic models}\label{sub-bioenergetic}
The dynamic energy budget (DEB) model provides the most complete and well-connected environmental variable effects for growth predictions by covering the entire life cycle of the fish \cite{Koo:12, LiS:08, MHGEJ:20}. 
The DEB model describes the metabolic processes of organisms in terms of energy. The description incorporates tools in the early life cycle, which enables obtaining beneficial knowledge before establishing new farms \cite{VLGTH:20, FGCG:14}, approximating fish production and the amount of food \cite{ChB:98}, and enhancing production by farming aquatic species in an integrated fashion \cite{RSPFG:12}. \textcolor{black}{In line with this}, the DEB growth model has become central in modeling and analyzing many fish species. This model can be formulated as the metabolic process, which is the difference between anabolism and catabolism, according to Ursin \cite{Urs:67}. 
\begin{enumerate}
\item \textcolor{black}{\noindent{\textbf{Individual fish growth models:}} The dynamics of the individual bioenergetic growth model are described in terms of fish biomass and population density, as follows \cite{CNMBL:21, Yan:98}}:
\textcolor{black}{\begin{equation}\label{sys1aa}
\!\!\!\!\!\!\!\!\frac{\der w}{\der t}= \underbrace{\Psi\big(f, T, DO\big)v(UIA)}_{\mbox{\scriptsize anabolism}} w^m - \underbrace{k(T)}_{\mbox{\scriptsize catabolism}} w^n,
\end{equation}}
where \textcolor{black}{$w$ denotes the individual fish weight}, $\Psi\big(f, T, DO\big)$ $(\si{kcal^{1\!-\!m}}\si{day^{-1}})$ and $ v(UIA)$ are the coefficients of anabolism, and $k(T)$ (\si{kcal^{1\!-\!n}}\si{day^{-1}}) represents the fasting catabolism coefficient formulated as follows:
\begin{equation*}\label{eq1a}
\!\!\Psi\big(f, T, DO\big)= h\rho f b(1-a)\tau(T)\sigma(DO),
\end{equation*}
and 
\begin{equation*}
k(T)=k_{\mbox{\tiny min}}\exp\Big({j(T-T_{\mbox{\tiny min}})}\Big).
\end{equation*}

The model constitutes the effects of water quality parameters, such as temperature ($T$), dissolved oxygen ($DO$), un-ionized ammonia ($UIA$), and feeding parameters, such as food availability \cite{Yan:98}. The relative feeding rate $f=\dfrac{r}{R}$ is expressed in terms of the maximal daily ration $R$ and the daily ration $r$. Table~\ref{para} summarizes the nomenclature and fish growth model parameters \cite{Yan:98}. The effects of temperature $\tau(T)$, un-ionized ammonia $v(UIA)$, fish mortality $k_1(UIA)$, and dissolved oxygen $\sigma(DO)$ on food consumption and their formulations are provided in the appendix \cite{Yan:98, CNMBL:21, Sink2010}.

Three assumptions were primarily considered while developing the \textcolor{black}{individual} fish growth model \eqref{sys1}:
\begin{itemize}
\item The proportionality coefficient of the potential net primary productivity to natural food assumed direct dependency on the standing crop of Nile tilapia and the relative feeding level. The potential net primary productivity is computed as the minimum of the total dissolved inorganic carbon, nitrogen, and phosphorus, assuming no threshold for the total dissolved inorganic nitrogen and phosphorus \cite{Yan:98, LiY:03}.

\item Due to the limited supply of nutrients (carbon, nitrogen, and phosphorus), the total production of phytoplankton as a natural food is proportional to the initial concentration of the limiting factor. Therefore, phytoplankton was assumed to be constant \cite{Yan:98, LiY:03}. 
\item The pH was constantly maintained. 
\end{itemize}

\item \textcolor{black}{\noindent{\textbf{Population dynamic fish growth models:}} Population models are crucial for describing potential pathways for dynamic recovery of the fish population and the patterns in biodiversity to understand and predict the future of marine resources. The fish growth population models based on the DEB integrate management variables, such as feeding, and environmental variables, such as temperature, dissolved oxygen, and ammonia, as effects for fish growth predictions and performance by covering the life cycle of the fish. Moreover, these models capture essential information on environmental and biological changes in fish, including mortality and fish stocking density \cite{Hiddink2008, Jenkins2020}.}
The dynamics of the bioenergetic growth models are described in terms of fish biomass and population density by \eqref{sys1}. The states $\xi$ and $p$ denote the total fish biomass and number of fish, respectively. In addition, $p_s$ and $\xi_i$ represent the stocking fish number and individual fish biomass during fish stocking, $k_1$ is the fish mortality coefficient, and $\xi_a$ denotes the mean fish biomass, which is equal to $\xi$ divided by $p$.
\begin{table*}[!t]
\begin{center}
\line(1,0){480}
\end{center}
{\normalsize
\begin{equation}\label{sys1}
\left\{\begin{array}{llllllll}
 \displaystyle \frac{\der \xi}{\der t}=\underbrace{p_s\xi_i}_{\mbox{\scriptsize fish stocking density}}+ \underbrace{\Psi\big(f, T, DO\big)v(UIA)}_{\mbox{\scriptsize anabolism}}\xi^m - \underbrace{k(T)}_{\mbox{\scriptsize catabolism}} \xi^n-\underbrace{pk_1(UIA)\xi_a}_{\mbox{\scriptsize fish mortality}}\\
\displaystyle \frac{\der p}{\der t}=p_s-\mbox{INT}(pk_1)
\end{array}\right.
\end{equation}
}
\begin{center}
\line(1,0){480}
\end{center}
\vspace{0.5cm}
\end{table*}


The fish growth model \eqref{sys1} can be written in a compact form as follows:
\begin{equation}\label{sys00}
\left\{\begin{array}{llllllll}
\displaystyle\frac{\der \xi}{\der t}= g\big(\xi,p \underbrace{f, T, DO}_{u},UIA, k_1\big)\\
  \displaystyle \frac{\der p}{\der t}=p_s-\mbox{INT}(pk_1)
\end{array}\right.
\end{equation}
where $\xi\in\mathbb{W}\subset\R$ denotes the state, and $u=[u_1,u_2,u_3]^T$ is the input vector. Additionally, $u\in\mathbb{U}\subset\R^3$ describes the manipulated control input vector corresponding to the feeding rate, temperature, and dissolved oxygen. The set of admissible input values $\mathbb{U}$ is compact. 
\end{enumerate}

\subsection{Challenges for building fish growth models}\label{Challenges-growth-model}
Modern dynamic systems based on the DEB are convincingly used to model fish growth. However, there are some goals and challenges associated with modeling and analyzing dynamic growth systems:

\noindent{\textbf{Growth state prediction:}}
In the DEB model for the fish growth case, we aim to predict the future growth state; however, long-term predictions may still be challenging. 

\noindent{\textbf{Design and optimization:}}
We can also tune growth model parameters to enhance performance and stability through design and optimization tools. 

\noindent{\textbf{Estimation and control:}}
It is possible to control the dynamic growth system using feedback from measurements of the growth system to actuate fish behavior. Thus, it is vital to have access to the full growth state from limited measurements.

\noindent{\textbf{Uncertainties:}}
There is uncertainty in the equations of fish growth, the output measurements (e.g., temperature, dissolved oxygen, and pH), and the specification of parameters (e.g., the coefficient of food consumption, coefficient of fasting, and parameters of food assimilation).

However, two primary challenges arise in building dynamic fish growth systems.
\begin{enumerate}
\def\labelenumi{\bf\theenumi)}\def\theenumi{\roman{enumi}}
\item {\bf Nonlinear growth models}:
Nonlinearity is one of the primary challenges in analyzing and controlling dynamic fish growth systems. Linear growth models derived, for instance, from the exponential growth at the initial stages can be fully characterized from the spectral decomposition perspective and generally lead to algorithms for estimation and control. However, this decomposition framework generally does not exist in nonlinear systems. Existing transformations of subspaces or approximations based on the ascending growth portion can provide local linearizations of the state around fixed points. However, the global analysis is still computationally expensive and squeezes the nonlinear prediction, estimation, and control theories of fish growth. Hence, developing a general framework for predicting, estimating, and controlling fish growth is a mathematical challenge in precision aquaculture.
\item {\bf Unknown dynamics:}
Although the DEB is convincingly used to model fish growth for many species, the lack of a unified governing growth equation remains a central challenge. With increasing environmental and management variable effects and complexity, deriving a growth model that encompasses these effects becomes challenging. Subsequently, the fish biomass state and some variables, such as fish mortality and coefficient of food consumption, can go unmeasured.
\end{enumerate}
Overcoming these two challenges can facilitate a better understanding of this DEB of the fish growth model.

\section{Traditional and optimal feeding and water quality control strategies}\label{Model-based}
\begin{figure}[!t]
\centering
   \begin{overpic}[scale=0.38]{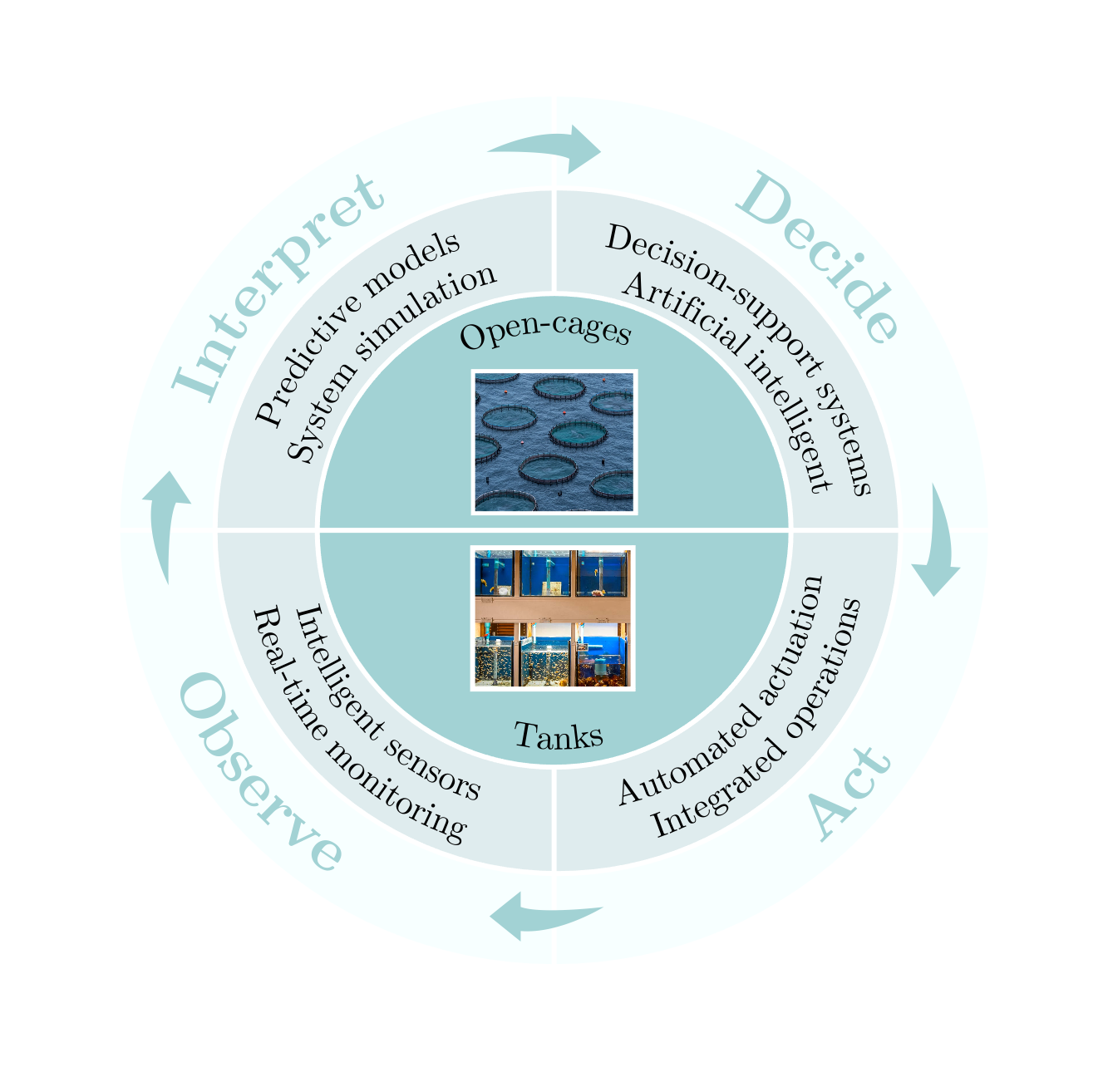}
   \end{overpic}\vspace{-0.7cm}
   \caption{Manual actions and monitoring, and experience-based
interpretation and decision-making.}\label{fig11ab} 
 \end{figure}

\begin{figure}[!t]
\centering
   \begin{overpic}[scale=0.45]{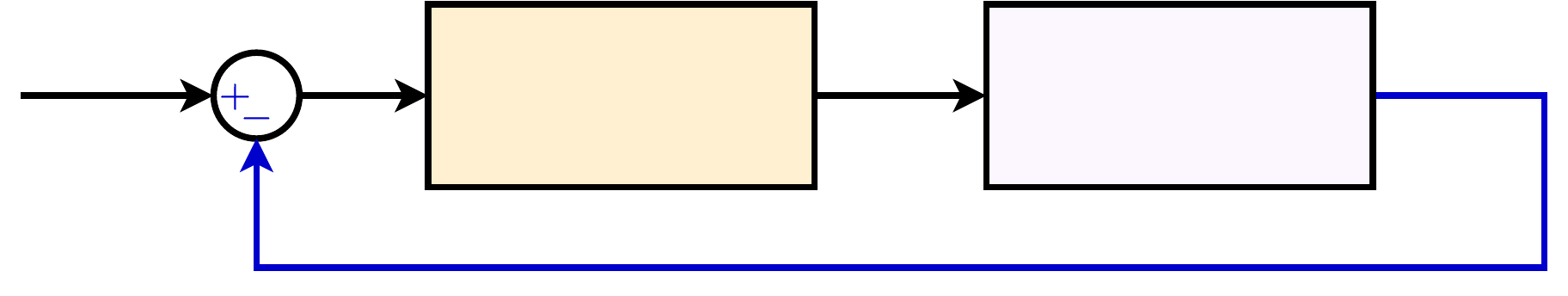}
   \end{overpic}\vspace{0.3cm}
  \put(-87,47){\small \textbf{System}}
  \put(-172,47){\small \textbf{Control}}
  \put(-129,-4){\small \textbf{Feedback loop}}
  \put(-168,34){\footnotesize Water quality}
  \put(-168,23){\footnotesize Feeding}
  \put(-79,28){\footnotesize Fish growth}
  \put(-191,34){\scriptsize error}
  \put(-110,34){\scriptsize input}
  \put(-27,34){\scriptsize output}
  \put(-235,34){\scriptsize \textbf {Desired}}
  \put(-235,20){\scriptsize \textbf{Reference}}\vspace{-0.25cm}
   \caption{Closed-loop control system scheme.}\label{Generalcontrolblock}
 \end{figure}
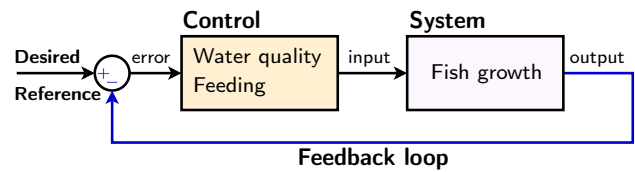

Developing control strategies in aquaculture was started to reduce labor costs and maintain a controlled environment. Today, most tasks in fish farming are conducted manually in different phases \cite{For:18}. These tasks are based on subjective experience and interpretation and represent manual versions of the feedback principles in control engineering, as illustrated in Figure~\ref{fig11ab}. Water quality and feeding are essential in aquaculture to enhance productivity and fish health. Providing more food than needed leads to water pollution and fish disease and stress \cite{AAKAFL:20}. Additionally, fish-feeding costs are roughly half the total production cost \cite{ASL:15, OVWB:20}. Hence, water quality and feeding are vital factors in managing the operational cost of aquaculture and influence fish health in a tank. Figure~\ref{Generalcontrolblock} illustrates the water and feeding feedback control loop derived from realizing the control system to monitor and achieve the desired fish growth status. 

The system/plant can be defined as a mathematical model. Models are generally categorized into physics-based, data-driven, and hybrid models. Physics-based models, also known as model-based, are derived from first principles. An example of a physics-based model is the governing fish growth equation that depends on the difference between anabolism and catabolism. Data-driven models, also known as model-free, are models generated from data. Such models depend on determining the correlations or mapping between input and output data. Hybrid models integrate model-based with model-free approaches, referring to the first principles and studying the correlation between input and output data. 

This section summarizes traditional and advanced feeding and water control strategies implemented on fish growth models.

\subsection{Traditional feeding and water controllers for monitoring fish growth}\label{tradi}
This section explores model-based feeding and water quality controllers for monitoring fish growth. We discuss classical controllers, such as bang-bang and proportional-integral-derivative (PID), and optimal and advanced control, such as MPC.

\subsubsection{Bang-bang control}\label{bang}
\begin{figure}[!h]
\vspace{0.5cm}
\centering
   \begin{overpic}[scale=0.34]{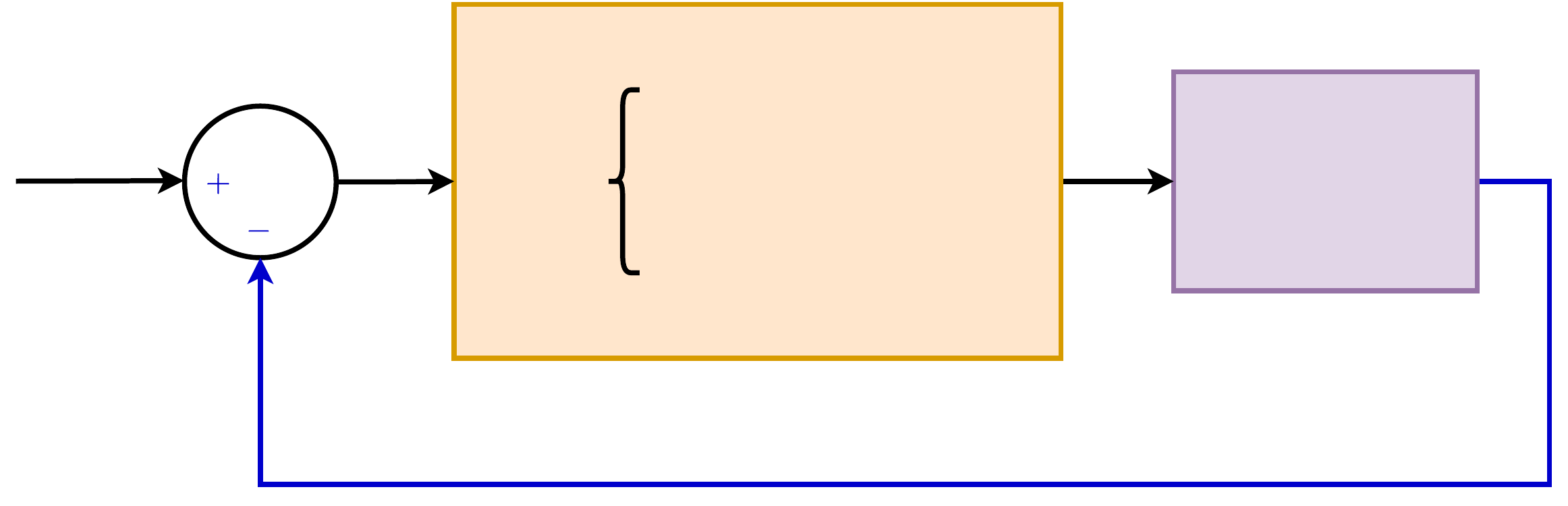}
   \end{overpic}
  \put(-135,55){\mbox{\scriptsize On, \ if} \scriptsize $e^j(t) >0$}
  \put(-135,35){\mbox{\scriptsize Off, \ if} \scriptsize $e^j(t) \leq0$}
  \put(-160,46){\scriptsize $u^j(t)=$}
  \put (-228,120){\scriptsize \textcolor{gray}{$j \ \ \ \ \ :$ \mbox{\scriptsize Number of control variables} }}
  \put (-228,110){\scriptsize \textcolor{gray}{$\mathcal{X}^j(t) :$ \mbox{\scriptsize Output measurements} }}
  \put (-228,100){\scriptsize \textcolor{gray}{$\mathcal{X}_d^j(t) :$ \mbox{\scriptsize Desired references} }}
  \put (-228,90){\scriptsize \textcolor{gray}{$e^j(t) \ =\mathcal{X}_d^j(t)-\mathcal{X}^j(t)$}}
  \put(-54,45){\scriptsize Fish growth}
  \put(-56,66){\scriptsize \textbf{System}}
  \put(-163,76){\scriptsize \textbf{Bang-bang control}}
  \put(-10,55){\scriptsize $\mathcal{X}^j(t)$}
  \put(-220,55){\scriptsize $\mathcal{X}_d^j(t)$}
  \put(-179,55){\scriptsize $e^j(t)$}
  \put(-72,55){\scriptsize $u^j(t)$}
   \caption{Closed-loop bang-bang scheme.}\label{Bang-Bangcontroller}
 \end{figure}
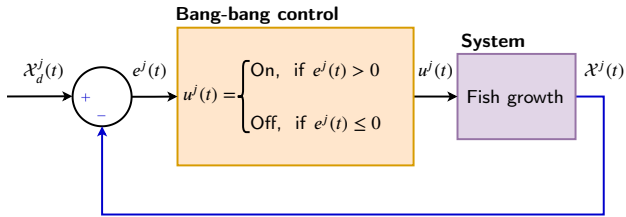

The bang-bang controller is a simple on-off device; yet, it is widely used in various market devices. It has been used to regulate water quality and automate feeding in aquaculture systems. In \cite{Francis2019}, the bang-bang controller was implemented to track the desired set-point of dissolved oxygen by switching the aerator on or off. An electric on-off actuator was implemented, and the objective was to switch the valves and pumps on or off depending on water quality measurements \cite{LEE1995205}. These controllers require a set of desired points for the controllers to meet. The mathematical representation for tracking the desired set-point is as follows: 
\begin{equation}\label{bang-bang cont}
u^j(t) = \left\{\begin{array}{lll}
 \displaystyle \mbox{On} \quad \quad \mbox{if} \ \ \ \ e^j(t) > 0,\\
\displaystyle \mbox{Off} \quad \quad \mbox{if} \ \ \ \ e^j(t)\leqslant 0,
\end{array}\right.
\end{equation}
where $u^j(t)$ denotes the input actions to the system, and $e^j(t)$ represents the error between the desired references and output measurements. For instance, if the temperature is set as the controllable parameter $T$, then the heater turns on if $T_{\mbox{\scriptsize desired}}- T> 0 \implies T<T_{\mbox{\scriptsize desired}}$ or turns off if $T_{\mbox{\scriptsize desired}} - T \leqslant 0 \implies T\geqslant T_{\mbox{\scriptsize desired}}$. The given conditions in the bang-bang controller are restricted to tracking desired references and can be set according to duty cycles, enabling an automated feeding system design. Figure~\ref{Bang-Bangcontroller} represents the implementation of the bang-bang controller.

\subsubsection{Proportional-integral-derivative controller}\label{pid}
A closed-loop or feedback system manipulates the input to obtain an output response close to a desired reference. Standard control systems comprise three primary blocks: the reference, controller, and system/plant. The reference block is the desired set-point or trajectory the user predefined so the output can track or follow. The controller block, such as a PID, evaluates the error between the desired reference and output measurement to provide an appropriate input action. The system/plant block can be considered a physical model that generates output from the input action. In the growth aquaculture models, feeding and water quality are manipulated variables because they affect the performance of the output. Similar to the desired reference, the controller can consider multiple variables to reduce the error between the outputs and desired references. 

The PID controller is calculated from the error feedback as follows:
\begin{equation}
u^j(t) = K_p^j e^j(t) + K_i^j \int e^j(t) \,dt + K_d^j \frac{d e^j(t)}{d t},
\end{equation}
where $u^j(t)$ denotes the input actions to the system, $j$ represents the number of control variables, $e^j(t)$ indicates the feedback or tracking error, and $K_p^j$, $K_i^j$, and $K_d^j$ are the proportional, integral, derivative gains, respectively. These gains are tunable parameters and require trials and tests to perform well. In addition, $K_p^j$ determines how far the outputs are from the desired references, and $K_i^j$ evaluates how long the outputs have been away from the desired references. Finally, $K_d^j$ indicates how fast the errors change.

\begin{figure}[!h]
\vspace{0.5cm}
\centering
   \begin{overpic}[scale=0.34]{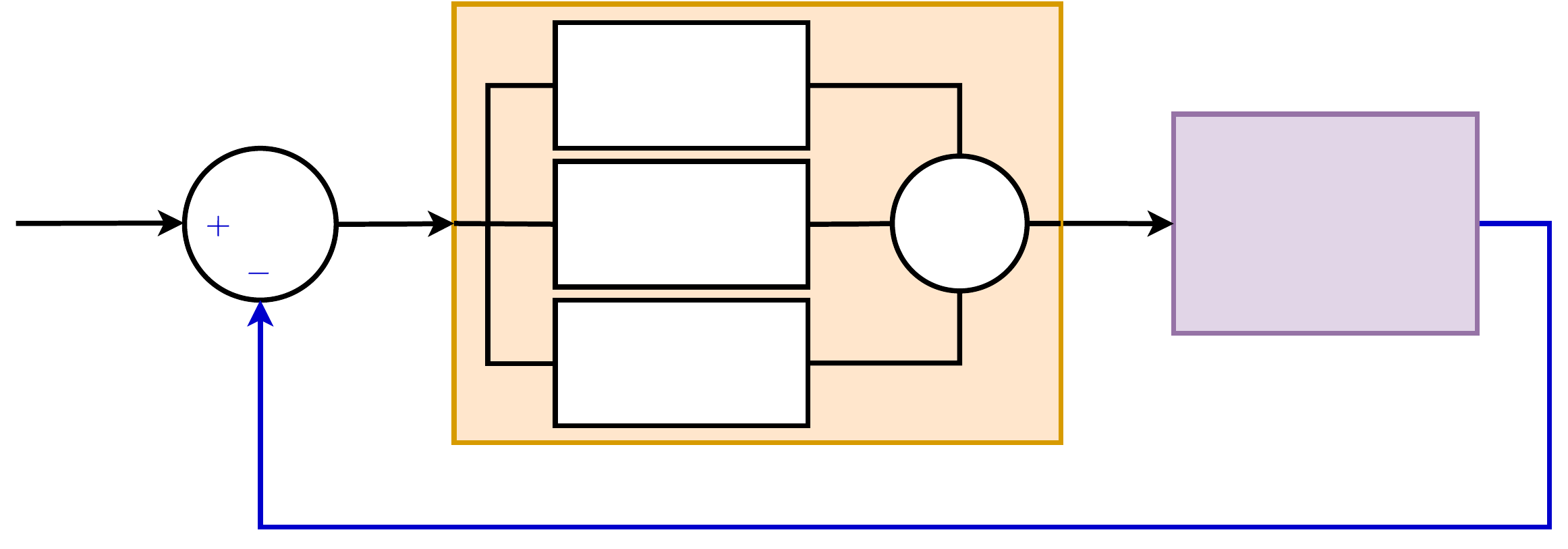}
   \end{overpic}
  \put(-141,65){\scriptsize $K_{p}^j \,e^j(t)$}
  \put(-147,45){\scriptsize $K_{i} \int e^j(t)\,dt$}
  \put(-145,25){\scriptsize $K_{d}^j \ \frac{d}{dt}e^j(t)$}
  \put (-228,120){\scriptsize \textcolor{gray}{$j \ \ \ \ \ :$ \mbox{\scriptsize Number of control variables} }}
  \put (-228,110){\scriptsize \textcolor{gray}{$\mathcal{X}^j(t) :$ \mbox{\scriptsize Output measurements} }}
  \put (-228,100){\scriptsize \textcolor{gray}{$\mathcal{X}_d^j(t) :$ \mbox{\scriptsize Desired references} }}
  \put (-228,90){\scriptsize \textcolor{gray}{$e^j(t) \ =\mathcal{X}_d^j(t)-\mathcal{X}^j(t)$}}
  \put(-54,45){\scriptsize Fish growth}
  \put(-91,45){\scriptsize $\Sigma$}
  \put(-56,66){\scriptsize \textbf{System}}
  \put(-163,83){\scriptsize \textbf{PID control}}
  \put(-10,55){\scriptsize $\mathcal{X}^j(t)$}
  \put(-220,55){\scriptsize $\mathcal{X}_d^j(t)$}
  \put(-179,55){\scriptsize $e^j(t)$}
  \put(-72,55){\scriptsize $u^j(t)$}
   \caption{Closed-loop proportional-integral-derivative (PID) control schematic.}\label{PID-controller}
 \end{figure}
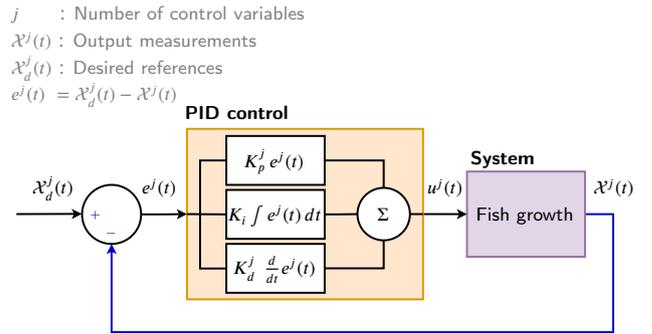

The PID controllers have been examined experimentally in aquaculture for water quality and feeding. For automatic feeding, the feeding rate and time have been implemented using PID controllers to enhance fish growth \cite{Zain2013}. In \cite{Pedro2020}, the nitrate concentration in water was controlled by a PID controller to track a desired reference. Figure~\ref{PID-controller} illustrates an example of a closed-loop control scheme that drives the measurements $\mathcal{X}^j(t)$ to their desired references. Assuming that the objective is to drive the temperature, dissolved oxygen, and feeding rate to the desired references, the measurements are $[\mathcal{X}^1(t), \mathcal{X}^2(t), \mathcal{X}^3(t)] = [T(t), DO(t), f(t)]$, and the references are $[\mathcal{X}_d^1(t), \mathcal{X}_d^2(t), \mathcal{X}_d^3(t)] = [T_d(t), DO_d(t), f_d(t)]$.

The design of closed-loop systems guarantees the best performance against requirements. However, traditional control, such as PID, may not be efficient for various fish growth tracking systems due to different management factors that account for the overfeeding specifications. In contrast, bang-bang control may induce undesirable responses, such as chattering, where the controllers switch on/off infinitely over a compact interval, and the state trajectories may not converge. Therefore, an optimal control law that ensures a tradeoff between optimal growth rate performance guarantees, cost minimization, and system complexity is relevant to increasing the overall performance.

\subsection{Optimal and advanced feeding and water controllers for optimizing fish growth}\label{advance}
Modern control relies on optimization techniques. Control and optimization are closely related problems in the behavioral domain. An optimization problem can be solved by minimizing the performance functional and using various stability criteria to design control laws. Optimal control approaches using optimization techniques were developed to enhance aquaculture process economics. Many of these optimal control techniques enhance the management and economics of aquaculture plants considering the best harvesting time and market variation \cite{optiz1}, \cite{Optz2}. However, previous studies have not examined all biological variables of the growth model and tend to focus on designing a generic framework. Another line of work is based on the fish price and mortality effects to provide the best feeding schedule \cite{optiz3}, \cite{Hea:95}. Therefore, there is an essential need to optimize feeding schedules and enhance fish farming productivity \cite{NGPKB:00}.

Furthermore, efficient control techniques are highly needed to monitor and control fish growth and react to different challenges associated with the practical aspects of aquaculture systems. Advanced control can have the most significant influence on modern aquaculture systems. For instance, the feeding control margins can be low to reduce overfeeding and, thus, wasted food that adversely affects water quality. 
Due to the feeding nature and environmental factors, most standard control methods cannot satisfy these constraints. Hence, the control problem can be formulated as an optimization problem, and the objective function is to target a desired fish growth under constraints, such as environmental factors, economic measures, and food amount \cite{CNMBL:21}. 

\subsubsection{Optimal control}\label{optimal}
In the case of the fish growth application, closed-loop PID-type and bang-bang controllers do not satisfy the desirable features, such as optimizing the feed conversion ratio by minimizing feeding while maximizing fish growth.
 \begin{figure}[!h]
\vspace{0.25cm}
\centering
   \begin{overpic}[scale=0.50]{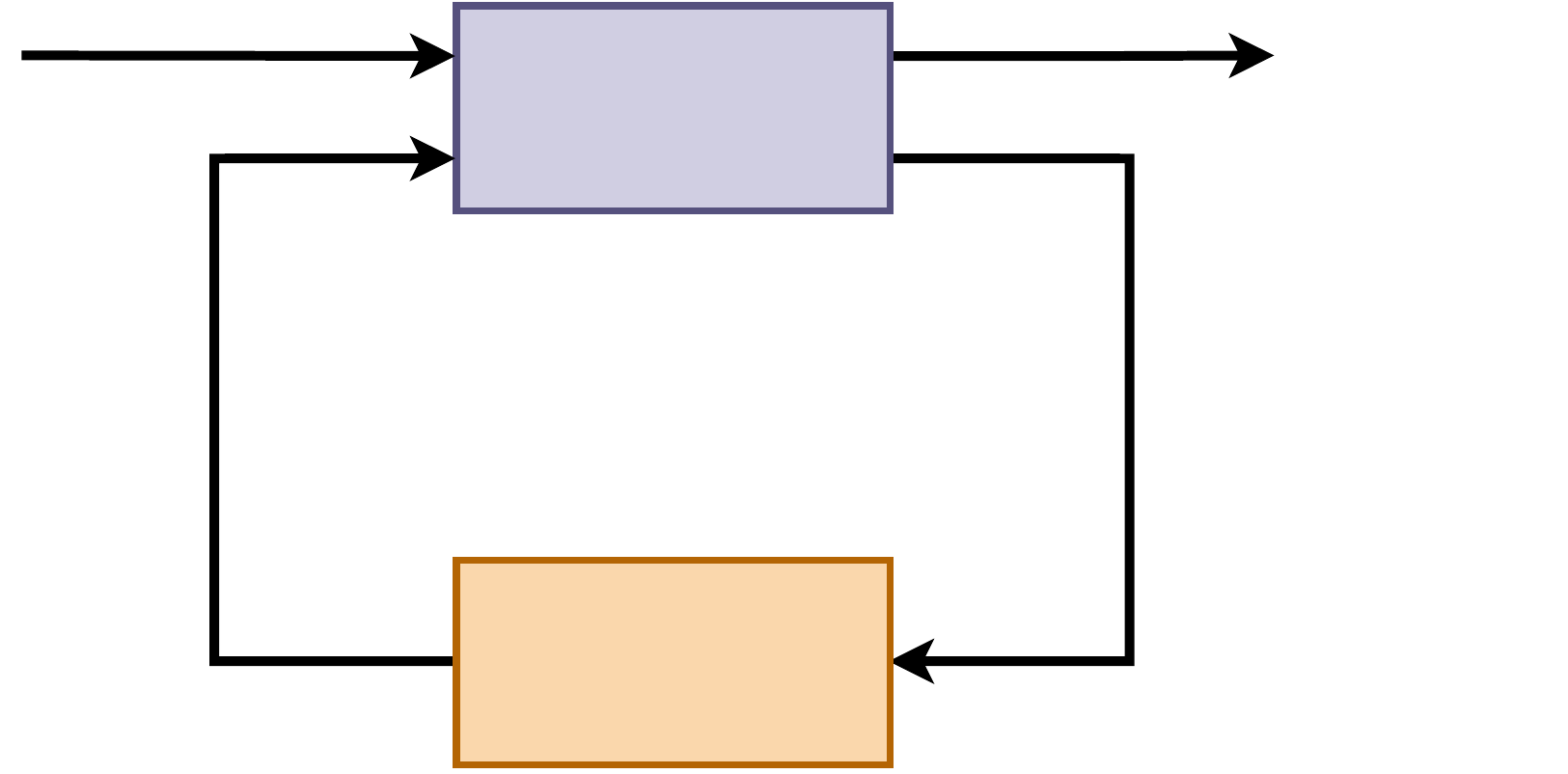}
   \end{overpic}\vspace{0.3cm}
  \put(-150,20){\small Optimal}
  \put(-148,7){\small control}
  \put(-158,98){\small {Fish growth}}
  \put(-90,95){\footnotesize Sensor}
  \put(-205,95){\footnotesize Actuator}
  \put(-80,113){\footnotesize Cost $J$}
  \put(-235,110){\footnotesize Desired reference}
  \put(-64,55){$y$}
  \put(-215,55){ $u$}
\vspace{-0.25cm}
   \caption{Optimal control scheme.}\label{opt}
 \end{figure}
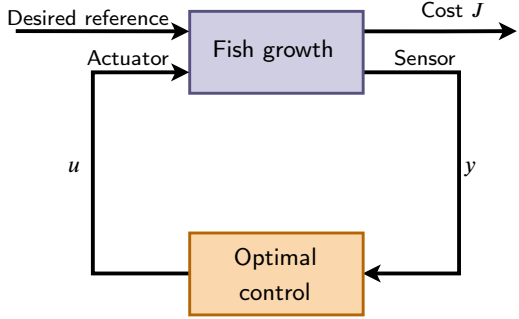
\vspace{0.5cm}
The optimal control problem for the growth rate is formulated as follows:
\begin{equation}\label{eq1-pf1}
 J(\tilde{w}(.), u(.))= \int_{t_0}^{t_f} \ell_{\tilde{w},u}\Big(\tilde{w}(.), u(.)\Big)\der \tau, 
 \end{equation}
where $t_0$ and $t_f$ denote the initial and final time that define the time horizon, respectively. Equation~\eqref{eq1-pf1} is subject to the system dynamics and constraint satisfaction. 
In addition, $\ell_{\tilde{w},u}\Big(\tilde{w}(.), u(.)\Big)$ is the performance functional (or stage cost), which is defined as a continuous, positive definite, and differentiable integrated function \cite{Lib:12}.
 
 \begin{remark}
For fish growth tracking control, it is not convenient to perform local linearizations on the growth model \eqref{sys00} due to the time-varying environmental factors of aquaculture, the nature of the desired trajectory, and the presence of model uncertainties and exogenous disturbances during system operation. Thus, a solution to adapting fish growth models using optimal control strategies can significantly improve feeding control performance while reducing the high computational cost.
\end{remark}

\subsubsection{Model predictive control}\label{mpc}
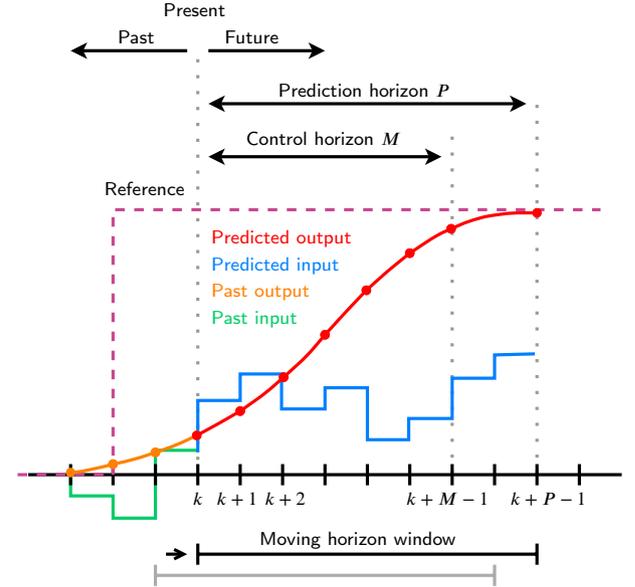
\begin{figure}[!h]
 \vspace{0.25cm}
\centering
   \begin{overpic}[scale=0.55]{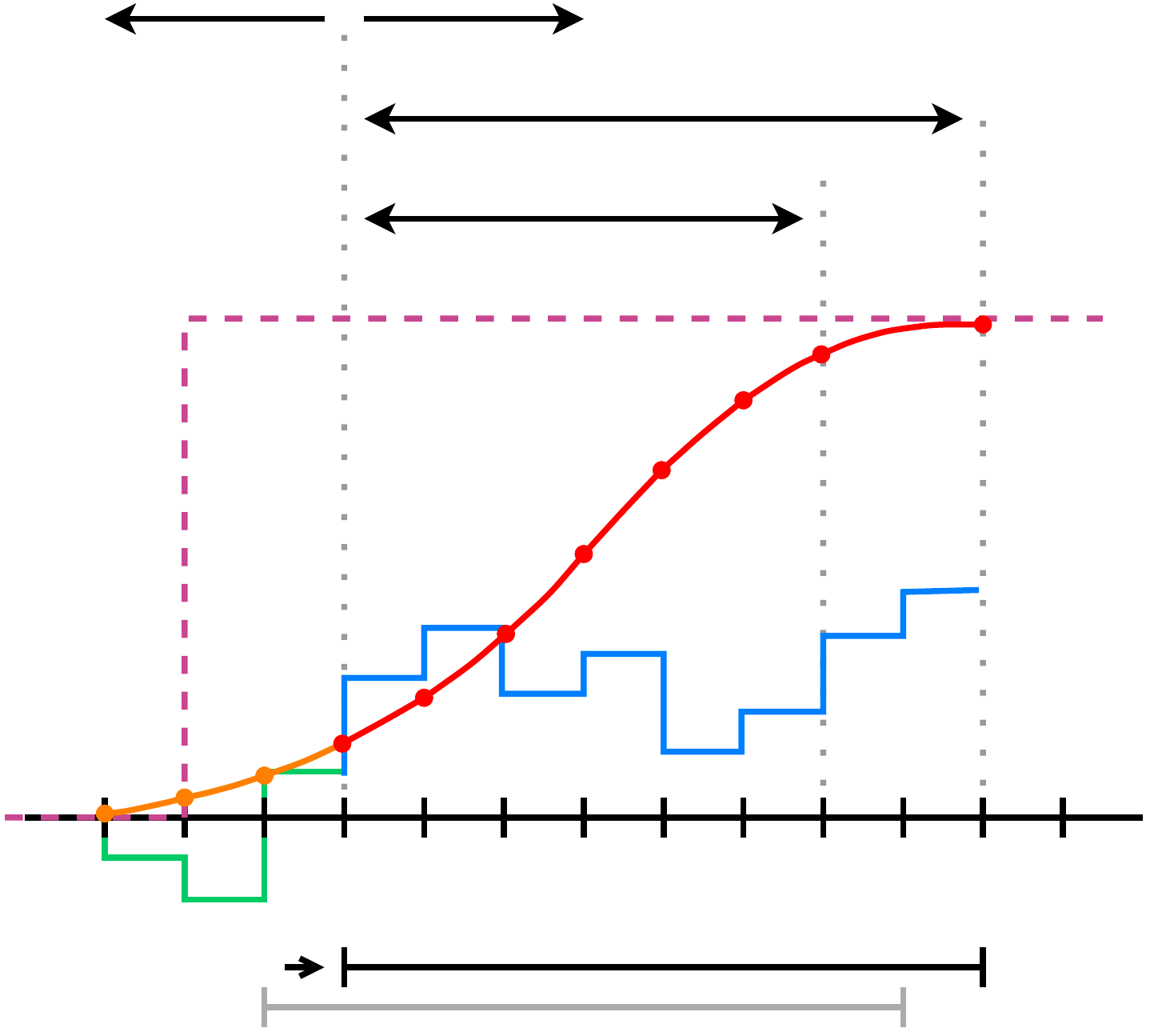}
   \end{overpic}
   \put(-142,167){\scriptsize Control horizon $M$}
   \put(-130,185){\scriptsize Prediction horizon $P$}
   \put(-150,205){\scriptsize Future}
   \put(-190,205){\scriptsize Past}
   \put(-173,215){\scriptsize Present}
   \put(-162,32){\scriptsize $k$}
   \put(-153,32){\scriptsize $k+1$}
   \put(-135,32){\scriptsize $k+2$}
   \put(-82,32){\scriptsize $k+M-1$}
   \put(-43,32){\scriptsize $k+P-1$}
   \put(-137,17){\scriptsize Moving horizon window}
   \put(-195,148){\scriptsize Reference}
   \put(-155,130){\scriptsize \textcolor[HTML]{FF0000}{{Predicted output}}}
   \put(-155,120){\scriptsize \textcolor[HTML]{007FFF}{{Predicted input}}}
   \put(-155,110){\scriptsize \textcolor[HTML]{FF8000}{{Past output}}}
   \put(-155,100){\scriptsize \textcolor[HTML]{00CC66}{{Past input}}}
   \vspace{-0.1cm}
   \caption{Model predictive control (MPC) framework.}\label{mpc2} 
 \end{figure}

The MPC has become the core of modern control and is operated in various industrial fields \cite{Raw:00, GPM:04}. It is used to find trajectories of nonlinear systems with constraints. Over a receding horizon, MPC aims to calculate the next control input action by solving an optimal control problem. This optimization process is repeated at each timestep to update the next control input \cite{KKB:18}, as illustrated in Figure~\ref{mpc2}.

\begin{figure}[!h]
\vspace{0.25cm}
\centering
\includegraphics[width=\linewidth]{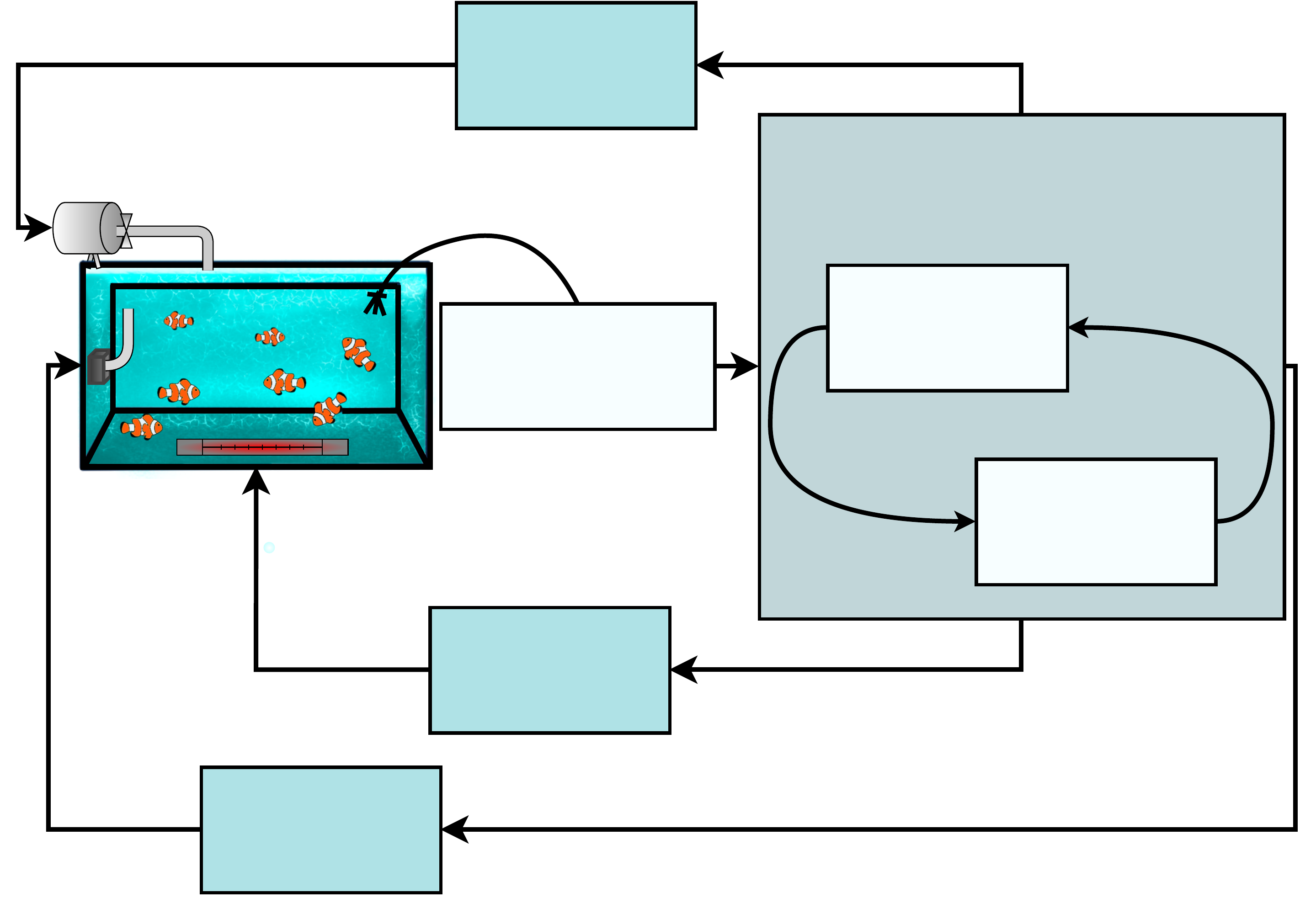}
\put(-147,152){\scriptsize Optimal}
\put(-146,142){\scriptsize feeding}
\put(-96,130){\footnotesize Model predictive control}
\put(-82,102){\scriptsize Optimizer}
\put(-50,65){\scriptsize Model}
\put(-38,105){\scriptsize predict}
\put(-98,62){\scriptsize control}
\put(-98,55){\scriptsize action}
\put(-156,98){\scriptsize Water quality}
\put(-156,89){\scriptsize measurements}
\put(-192,15){\scriptsize Aerator}
\put(-189,7){\scriptsize pump}
\put(-150,38){\scriptsize Heater}
\caption{Model predictive control framework to optimize factors strongly influencing fish growth, such as feeding rate, dissolved oxygen, and temperature.}\label{fig_MPC}
\end{figure}
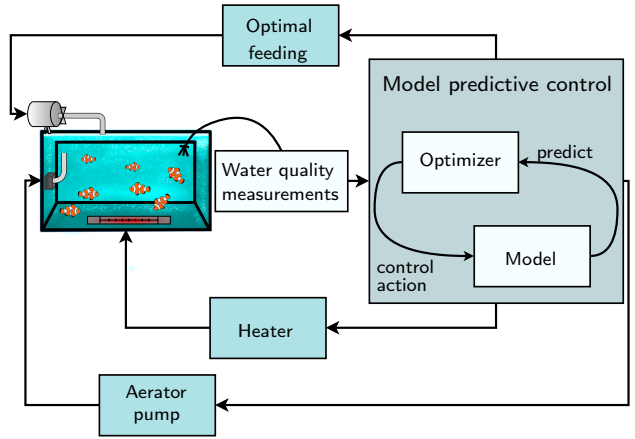

The first work using MPC-based on the receding-horizon framework for the reference growth tracking problem in the aquaculture system was \cite{CNMBL:21}, which maximizes fish biomass production while minimizing production costs under a representative bioenergetic growth model, as illustrated in Figure~\ref{fig_MPC}. 
In \cite{CNMBL:21}, the reference tracking problem of fish growth is formulated as a minimization with a finite-time prediction horizon as follows: 
 \begin{subeqnarray}\label{eq1-mpc1}
&&\!\!\!\!\!\!\!\!\!\!\!\!\!\!\!\!\displaystyle \min_{{u \in \mathcal{U}(\varepsilon)}}\!\!J\!=\!\!\int_{t_k}^{t_{k+N}}\!\!\ell(.) \der \tau\!\!\!\! \slabel{eq-mpc1a}\\ 
&&\!\!\!\!\!\mbox{s.t}\quad \dot{\tilde{w}}(t)= g\big(\tilde{w}(t),u(t)\big) \slabel{eq-mpc1b}\\
&&\!\!\!\!\! u_{\mbox{\scriptsize min}}\leqslant u(t) \leqslant u_{\mbox{\scriptsize max}}, \quad \forall t \in[t_k,\, t_{k+N}]\slabel{eq-mpc1c}\\
&&\!\!\!\!\! \Delta u(t_k)= u(t_k)-u(t_{k-1})\slabel{eq-mpc1cbis}\\
 & &\!\!\!\!\! w_0 \leqslant \tilde{w}(t) \leqslant w_{\mbox{\scriptsize end}} , \quad \forall t \in[t_k,\, t_{k+N}]\slabel{eq-mpc1c0} \\
&&\!\!\!\!\! \tilde{w}(t_k)=w(t_k), \quad \tilde{w}(0) =w(t_0) \slabel{eq-mpc1d},
  \end{subeqnarray}
where $N$ refers to the prediction horizon, $\tilde{w}$ represents the predicted trajectory of the fish growth over $[t_k,\, t_{k+N}]$, and $w(t_k)$ denotes the fish biomass state measurement obtained at time $t_k$. The stage cost $\ell(.)$ can represent an economic measure or tracking error performance of the fish growth rate (see \cite{CNMBL:21} for a more detailed treatment of the stage cost). Finally, $w_0$ and $w_{\mbox{\scriptsize end}}$ are the constraints of the desired initial and maximal fish weight.

From known governing DEB equations (i.e., physical laws) of the growth that arises from first principles, model-based control approaches enable deriving appropriate controllers to ensure control requirements and specifications. However, the resulting control performance guarantees depend on the precision of the fish growth models, and there is no improvement once the control-designed procedure is established. In contrast, the growth model \eqref{sys00} presents a high-dimensionality challenge associated with the growing population dynamics.

\section{Model-free-based reinforcement learning control}\label{Model-free}
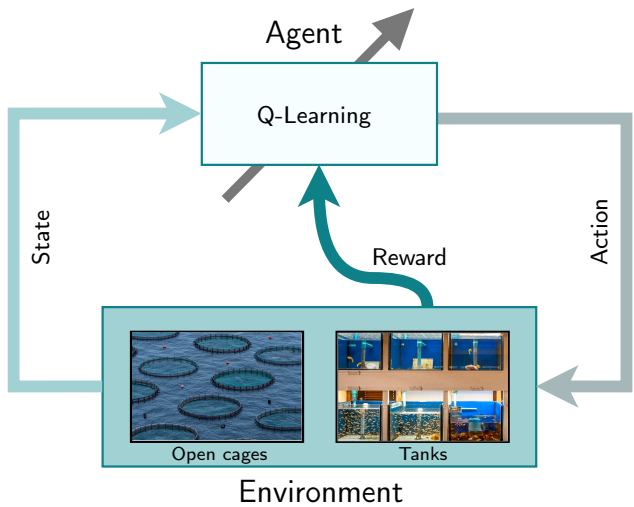
\begin{figure}[!h]
\vspace{0.25cm}
\centering
   \begin{overpic}[width=\linewidth]{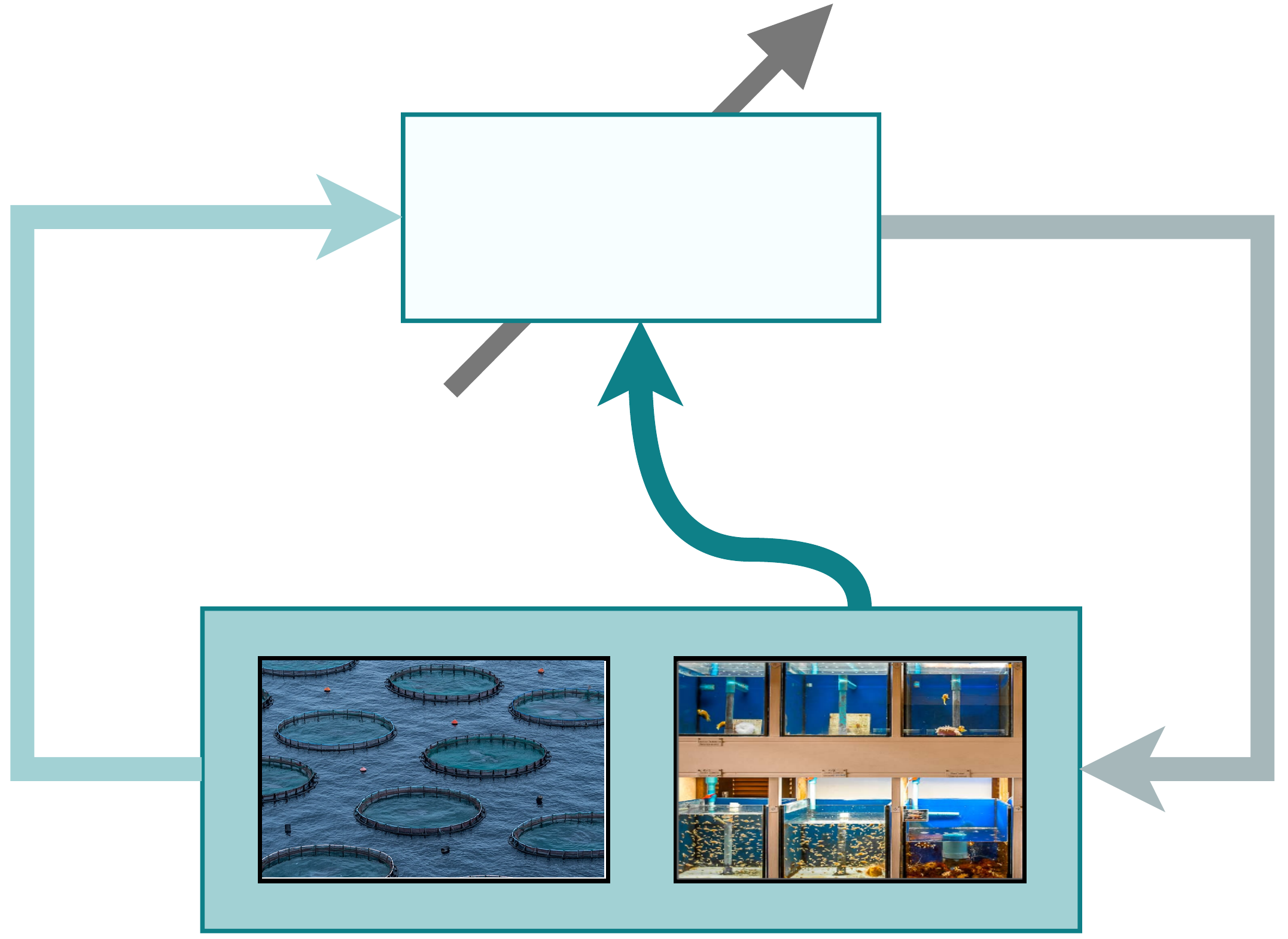}
   \end{overpic}\vspace{0.1cm}
   \put(-143,130){\small Q-Learning}
   \put(-140,160){\large Agent}
   \put(-150,-12){\large Environment}
   \put(-227,77){\small \rotatebox{90}{State}}
   \put(-18,77){\small \rotatebox{90}{Action}}
   \put(-175,4){\scriptsize {Open cages}}
   \put(-90,4){\scriptsize {Tanks}}
   \put(-100,77){\small Reward}
   \caption{Reinforcement learning structure.}\label{RLplot} 
 \end{figure}

The model-free control of complex systems enables learning control laws. Specific model-free control methods include genetic algorithms, adaptive neural networks, and RL \cite{FlP:02}, an essential discipline at the intersection of control and machine learning. The most common framework for RL is the Markov decision process (MDP), where the dynamics and control are described in a probabilistic setting. Figure~\ref{RLplot} illustrates the RL structure.

The RL methods learn the optimal control policy in an interactive environment, which does not require knowing the exact dynamic model. The RL method aims to take action in an environment to maximize the final reward, and RL-based control can be a practical alternative to classical control methods because it does not require an exact dynamic model to solve for the controller. From interaction with the real environment, the optimal action can be derived. Further, RL has been implemented in control applications, including set-point tracking error control problems \cite{LeL:05}, position tracking control and suspension regulation \cite{WSYW:18, CDWMPA:18, ZYSZT:21}, robust quadrotor control \cite{WSHS:20}, and the process industry under a changing environment \cite{LiD:20}.

Most model-based control approaches find the fish growth tracking problem challenging due to interactions between multiple inputs and nonlinear couplings, such as dissolved oxygen, temperature, un-ionized ammonia, and the fish growth model uncertainty. Additionally, the fish growth rate varies in practice and cannot be easily estimated due to the complex aquaculture conditions and time-varying environmental factors. With the development of machine-learning techniques, the problem of tracking the fish growth trajectory has been developed as sampled-data optimal control using the MDP with finite state-action pairs on the simulated fish growth trajectory data to sufficiently emulate the aquaculture environment \cite{CNMBL:22}.

Recently, deep learning technology has demonstrated the ability to recognize the wave size to determine whether to continue or stop through an automatic fish-feeding system \cite{HCHL:22} and has developed an aquaculture simulator to facilitate the feeding control process \cite{KII:20}. However, in aquaculture, RL is primarily applied to fish robotic research and schooling navigation areas \cite{YCWTZ:21, YWYYZ:21, VNK:18}. To this end, few studies have assessed fish-feeding control based on the RL methodology in monitoring and optimizing fish growth rate production for aquaculture systems. This scarcity motivated the development of Q-learning-based algorithms that learn the optimal feeding control policy for the fish growth rate for aquaculture in our recent work \cite{CNMBL:22}.


 
 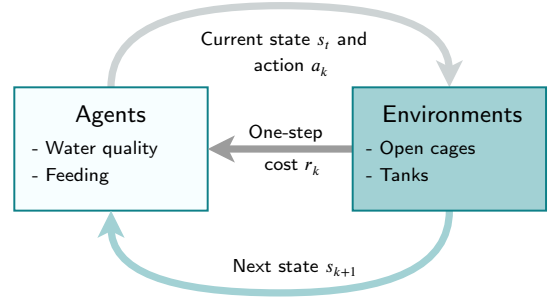
\begin{figure}[!t]
\centering
   \begin{overpic}[scale=0.34]{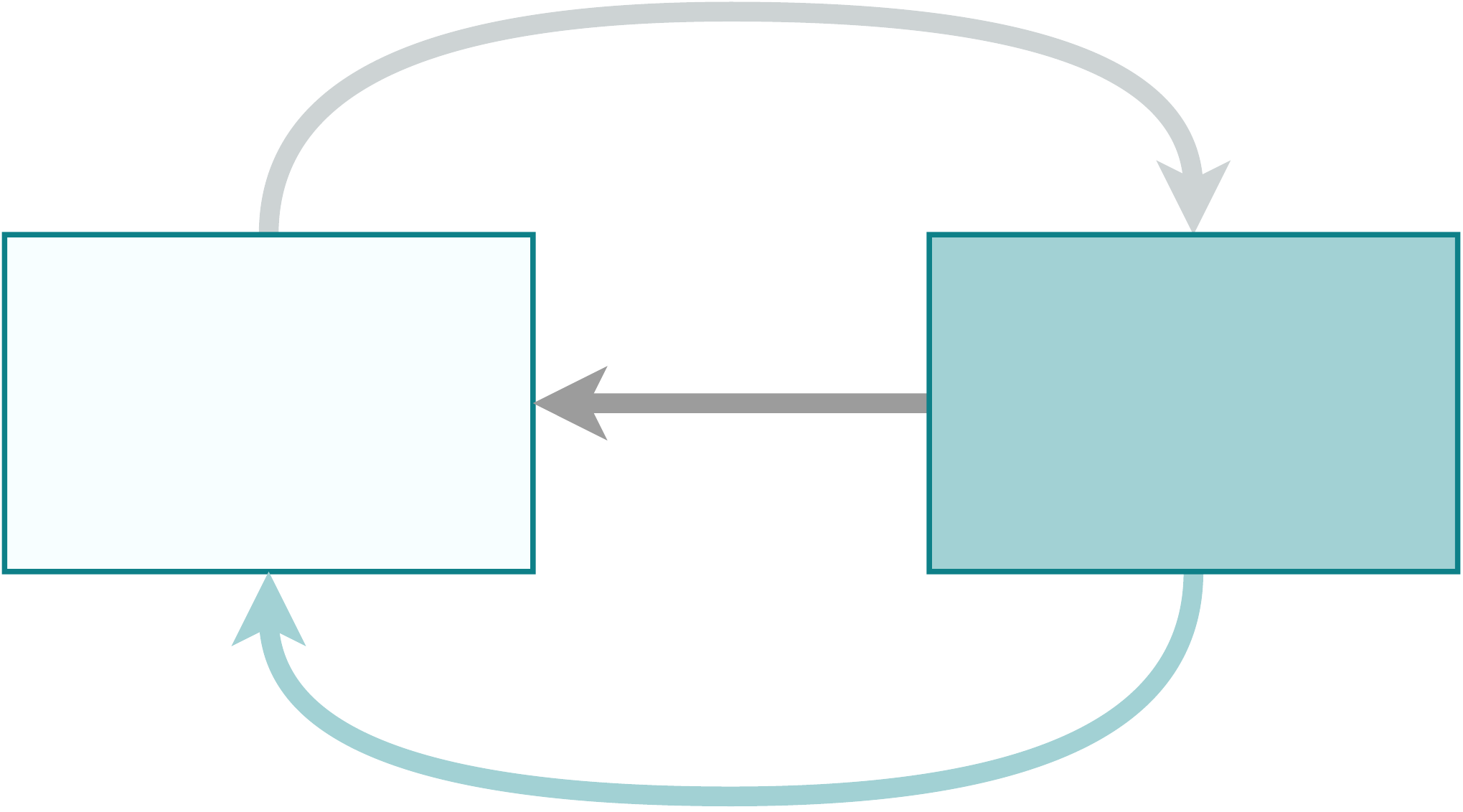}
   \end{overpic}
   \put(-177,66){\small Agents}
   \put(-193,54){\scriptsize - Water quality}
   \put(-193,44){\scriptsize - Feeding}
   \put(-62,66){\small Environments}
   \put(-68,54){\scriptsize - Open cages}
   \put(-68,44){\scriptsize - Tanks}
   \put(-130,95){\scriptsize Current state $s_t$ and}
   \put(-110,85){\scriptsize action $a_k$}
   \put(-112,60){\scriptsize One-step}
   \put(-106,48){\scriptsize cost $r_k$}
   \put(-118,10){\scriptsize Next state $s_{k+1}$}
   
   \caption{Markov decision process (MDP) framework.}\label{MDP-plot} 
 \end{figure}

\subsection{Markov decision process}\label{sec-MDP1}
Learning decisions that drive deterministic and probabilistic rewards usually employ the MDP \citep{Put:14}. The RL framework guarantees the Markov property, which considers that the current state of an agent takes all relevant information. We define the MDP model by the following components: i) a state space $\mathcal{S}$, ii) an action space $\mathcal{A_s}\subset \mathbb{R^+}$, iii) a one-step cost function $r(\textbf{s},\textbf{a}): \mathcal{S} \times \mathcal{A_s} \rightarrow \mathbb{R^+}$, and iv) a one-step transition probability $p(\textbf{s}_{k}\!\mid\!\textbf{s}_1,\textbf{a}_1,\cdots,\textbf{s}_{k-1},\textbf{a}_{k-1})$ in the stochastic case or a transition function of the ground truth environment $h$ in the deterministic case. 

In the MDP framework, the agent takes action $\textbf{a}_k$ in the environment from a given state $\textbf{s}_k$. Then, the environment outputs a new state $\textbf{s}_{k+1}$ and a reward $r_k = r(\textbf{s}_k, \textbf{a}_k)$, which can indicate the performance of the taken action in the environment from a given state as illustrated in Figure~\ref{MDP-plot}. The Q-learning goal based on the MDP is to derive an optimal policy $\pi$ to maximize the long-term cumulative cost. Hence, the optimization strategy is formulated as follows \cite{Ber:96,sutton1998r, Ber:05}:
\begin{equation}\label{reward1a}
\max_{\pi\in{\cal R}} \hat{J}(\pi)= \max_{\pi\in\cal R} \sum_{k=1}^N\gamma^{k-1}r_k\!\mid\!\pi,
\end{equation}
where ${\cal R}$ is the policy space, and $\gamma$ represents a discount factor with $0\!<\!\gamma\!<\!1$.

\subsection{Markov decision process for the fish growth problem}\label{sec-MDP}
The MDP is used through the framework of Q-learning to describe the aquaculture environment and construct the RL environment where the Markov property is assumed to hold. The current state and action include the needed information to determine the next state \cite{sutton1998r, BeT:96, watkins1992q, Bel:03, Ber:05, Pow:07, Sug:15}. The fish growth problem is defined by the finite MDP as follows:\\
\noindent{\textbf{State}:} The set $\mathcal{S}$ comprises finite states that contain all possible configurations of the aquaculture system, such as the weight, number, or age of the fish.\\
\noindent{\textbf{Action:}} The set $\textbf{a} \in \mathcal{A_s}$ comprises the possible allowable actions that drive the current state $\textbf{s}$ to the next state $\textbf{s'}$. The finite space of action is expressed as $\mathcal{A_s}\subset \mathbb{R^+}$. Fish farmers usually use two systems to grow fish: open cages or tanks (see Figure~\ref{sensing-monitoring}). Consequently, the fish growth model \eqref{sys1} describes the transition from the current state to the next state. Hence, the agent follows a deterministic policy for any trajectory generated in the state-action space.\\
\noindent{\textbf{Reward:}} The variable $r(\textbf{s},\textbf{s'})$ is reward obtained after taking action $\textbf{a}$ in the environment from state $\textbf{s}$ to the next state $\textbf{s'}$. The fish growth problem aims to maximize fish growth to a desired weight with minimum feed ration $\big(f\big)$ in an open-cage or tank aquaculture system, where the temperature is controlled. For both environments, the formulation of the reward function $r\Big(\textbf{s},\textbf{a}\Big)$ is as follows:
\begin{equation}\label{reward2a}
r_k\Big(\textbf{s}_k,\textbf{a}_k\Big)= \bar{\ell}(.),
\end{equation}
where $\bar{\ell}(.)$ denotes an economic measure or tracking error performance of the fish growth rate (see \cite{CNMBL:21, CNMBL:22} for the choice of the reward cost).

This reward \eqref{reward2a} maximizes the fish growth and penalizes the feed ration under the optimal temperature profile for aquaculture tank systems. For open cages, the reward maximizes fish growth to a specified target and minimizes the feed ration.

\subsection{Q-learning algorithm-based temporal difference method}\label{sec-Q}
Q-learning is an advanced value-based RL approach that seeks the optimal policy \citep{watkins1992q}. In particular, it learns the optimal policy without requiring complete system knowledge \eqref{sys00}. The Q-learning function $Q(\textbf{s},\textbf{a})$ expresses all allowable actions $\textbf{a}$ at given states $\textbf{s}$. At each sampling step $k$, the weights interact with the aquaculture environment, and the temporal difference updates the Q-learning function after using the sampling experiences. The Q-learning function is 
\begin{align}\label{QRL}
&\!\!\!\!\!\!\!\!\!Q\Big(\textbf{s}_{k},\textbf{a}_{k}\Big)\!\leftarrow\!Q\Big(\textbf{s}_{k},\textbf{a}_{k}\Big) \notag \\ &+\alpha\left[ r_k\Big(\textbf{s}_k,\textbf{a}_k\Big)\!+\!\gamma \max_{\textbf{a}} Q\Big(\textbf{s}_{k+1},\textbf{a}_{k+1}\Big)\!-\!Q\Big(\textbf{s}_{k},\textbf{a}_k\Big)\right],
\end{align}
where $Q\Big(\textbf{s}_{k},\textbf{a}_{k}\Big)$ is the value function of the state-action pair $\Big(\textbf{s}_k,\textbf{a}_k\Big)$ at each step $k$, and $r_k\Big(\textbf{s}_k,\textbf{a}_k\Big)$ represents the reward function. In addition, $\alpha$ is the learning rate, and $\gamma$ is the discount factor.

The Q-learning algorithm uses the exploration and exploitation frameworks to ensure a globally optimal policy $\pi^*$. Excessive exploration is needed to learn the generated data thoroughly in the implementation phase. Thus, an obtained suboptimal policy initializes the on-policy learning for real experiments. Further, a greedy parameter $\epsilon$ that decays exponentially with the increase in training episodes $i$ is introduced to alternate from the exploration to the exploitation phases as follows:
\begin{equation}\label{greddy_annealing}
\epsilon = 1-\epsilon_0 \exp\Big(\dfrac{i}{t_\epsilon} \Big),
\end{equation}
where $t_\epsilon$ refers to the duration of the exploration phase and the starting point of the exploitation phase, and $i$ defines the current training episode. Algorithm \ref{algo1} presents the pseudocode to implement an off-policy iteration algorithm with a greedy action.
		\begin{algorithm}[htbp]
          	\removelatexerror
        	\SetAlgoLined
        	\SetKwInOut{Input}{input}\SetKwInOut{Output}{output}
			\Input{$\mathcal{A}$: set of possible actions\\$\mathcal{S}$: set of possible growth states\\ $\xi^d$: target fish weight\\ $\xi_0$: initial fish weight\\ $\epsilon_0$: initial greedy policy (exploration factor)\\ $t_\epsilon$: exploration phase duration}
			\Output{$\pi^*$ optimal control policy }
			$\diamond$ Initialize policy\;
			$Q(\textbf{s},\textbf{a})\leftarrow 0$, \quad $\pi=\pi_Q$\quad ($\pi_Q$ is derived from $Q$)\;
			\While{stop = 0}{$\diamond$ Policy improvement\;
			$\textbf{s}\leftarrow 0$, \quad $\pi_{old}=\pi$\;
			\While{ $s \neq$ `Terminal'}{
			    $\checkmark$ compute $\epsilon$ from \eqref{greddy_annealing}\;
			    $\checkmark$ choose an action $\textbf{a}$ from state $\textbf{s}$ using $\pi$ ($\epsilon$- greedy)\;
			    $\checkmark$ observe the new state $\textbf{s'}$ and reward $r$\;
			    $\checkmark$ $\textbf{s}\leftarrow \textbf{s'}$, \quad $episode \leftarrow ~episode+1$\;
			    $\diamond$ Update Q-table and policy $\pi_Q$\;
				$\delta=r\big(\textbf{s},\textbf{a}\big)+\gamma \max_{\textbf{a}} Q\big(\textbf{s'},\textbf{a'}\big)\!-\!Q\big(\textbf{s},\textbf{a}\big)$\; 
				$Q\big(\textbf{s},\textbf{a}\big) \leftarrow Q\big(\textbf{s},\textbf{a}\big)+\alpha \delta$\;
				$\pi =\pi_Q$\;
				$\diamond$ check the stopping criteria after each episode\;
				\If{ $\pi_{old}=\pi$}{ $stop \leftarrow 1$}
			}
			$\pi^* \leftarrow \pi$
			}
			\caption{\small{Policy iteration based on Q-learning~\cite{CNMBL:22}}}\label{algo1}
		\end{algorithm}

%

\section{Synergetic model-based and model-free control concepts}\label{Integrated-model}
Model-based control, such as MPC, and model-free control, such as RL, are well-known methods for optimal control. The control theory community has studied MPC, whereas the machine-learning community has promoted RL. The MPC method is recognized for its robustness, feasibility, and stability and can handle constraints. However, it is not an adaptable method and has a computational burden in the online case. In contrast, RL is known for its adaptability and low online complexity. Thus, it struggles to handle constraints. 

Despite the drawbacks of the two methods, they share some similarities and can complement each other. The relationship between MPC and RL was first recognized in \cite{Sutton1992}. Since then, this relationship has been an attractive area bridging the control theory and machine-learning communities \cite{Daniel2017}. Nevertheless, a few studies have combined model-based and model-free approaches. Recent research that combined both concepts has presented impressive results in robotics \cite{Williams2017, Greatwood2019} and energy management \cite{Javier2022}. 

\subsection{Learning MPC cost from RL}
The synergy between the two optimal control strategies, MPC and RL, has recently shown promising benefits in satisfying constraints through several applications. In \cite{Daniel2017, Hoeller2020}, the relationship of combining MPC with RL is observed in the objective function $J$ in MPC. In particular, the knowledge of policy and value function structures can be exploited from the cost of MPC. An actor-critic structure can solve the optimal value function due to its suitability to combine MPC as an actor agent. This combination is beneficial to reduce the off-line training complexity and enhance the convergence in RL. 

However, one of the drawbacks of this approach is that how to use economic measures in the cost function of the MPC is unclear. The objective function depends heavily on the value function, which reduces the complexity of RL. In contrast, how to add economic measures to the MPC objective function is not obvious. An off-policy learning-based MPC method called the guided policy search, which takes advantage of model-based methods at training time, was proposed in \cite{ZKLA:16}. The guided policy search iteratively collects training data using MPC to convert the policy \cite{KHLZ:19, OSB:16, RoB:20}. 

Despite the overall potential features of learning the MPC cost from RL and the ability of model-based controllers, designing such a system remains a significant challenge. Further, these approaches learn black-box control policies that undergo rough generalizations. Recently, a novel algorithm called reinforced predictive control (RL-MPC) that merges the relative features was proposed in \cite{Javier2022}. The new RL-MPC algorithm enables continuous learning and meets the required constraints. Furthermore, it can achieve performance levels comparable to MPC \cite{Javier2022}. Additionally, the RL-MPC algorithm possesses relevant features and can be a good candidate for assessing fish growth feeding and water quality controls.

To the best of our knowledge, there are no results of combining or merging model-based and model-free controllers to optimize fish growth and survival in aquaculture. Inspired by the previous study on RL-MPC proposed in \cite{Javier2022}, this section addresses a novel algorithm called reinforced MPC to assess fish growth in aquaculture. 

\begin{table*}[t]
\caption{Properties of learning model predictive control (MPC) cost from reinforcement learning (RL) and learning the RL rewards from MPC.}
\begin{center}
\begin{tabular}{| c || c c |}
 \hline
 ~~~~~~~~~~~~\textbf{Property}~~~~~~~~~~~~ & ~~~~~~~~~\textbf{Learning MPC cost from RL}~~~~~~~~~& ~~~~~~~~~\textbf{Learning the RL rewards from MPC}~~~~~~~~~ \\ 
 ~~~~~~~~~~~~~~~~~~~~~~~ & ~\cite{Daniel2017,Hoeller2020}~~~~~& ~~~\cite{Javier2022}~~~~~~~ \\ \hline
System model & Requires a model \ding{55} & Requires an identified/calibrated model \ding{51} \\ 
RL reward & Depends on LQR structure \ding{55} & Depends on the MPC cost \ding{51}\\ 
\multirow{2}{*}{Linearity} & Requires a linear system \cite{Daniel2017} \ding{55}& \multirow{2}{*}{Does not require a linear system \ding{51}} \\
& ~~~~Does not require a linear system \cite{Hoeller2020} \ding{51}~~~~ & \\
Objective function (MPC) &Depends on the value function \ding{55}& Depends on the economic measures \ding{51} \\ 
Constraints & Satisfied from MPC \ding{51} & Satisfied from MPC \ding{51} \\ 
Stability & Satisfied from LQR \ding{51} & ~~~Not rigorously proven (relies on MPC only) \ding{55}~~~ \\ 
Feasibility & Satisfied from LQR \ding{51}& Not rigorously proven (relies on MPC only) \ding{55} \\ \hline
\end{tabular}
\end{center}
\label{sample-table}
\end{table*}

\subsection{Learning the RL reward from MPC}
This section aims to learn from the environment while imposing constraint conditions. The link between MPC and RL appears in terms of the reward function $r_{k+1}$ and discount factor $\gamma$ as follows \cite{Javier2022}: 
\begin{equation}\label{reward7}
r_{k+1}\Big(\textbf{s}_{k+1},\textbf{a}_{k+1}\Big)= -(J_{k+1} - J_k),
\end{equation}
where $J_k$ and $J_{k+1}$ are the objective functions of the MPC at time steps $t_{k}$ and $t_{k+1}$, respectively. Usually, the reward function is formulated as a maximization function where the objective cost of the model-based approach is considered a minimization problem. The appearance of the negative sign in \eqref{reward7} is because of the synergy between the two models. Similarly, the discounted factor $\gamma$ is structured in terms of the moving and control horizons as in \eqref{discount-factor}: 
\begin{equation}\label{discount-factor}
\qquad\qquad\qquad\gamma= 1-\frac{M}{N},
\end{equation}
where $M$ denotes the control horizon, and $N$ represents the prediction horizon from MPC, as illustrated in Figure~\ref{mpc2}. 
\begin{remark}
The discount factor $\gamma$ determines how the agent learns from its reward. When $\gamma =0$ or $M=N$, the agent learns only about actions that result in an immediate reward. In contrast, when $\gamma \approx 1$ or $M>>N$, the agent learns from the sum of the future rewards. Therefore, the choice of the control and prediction horizons in \eqref{discount-factor} determines the agent's learning process. 
\end{remark}

The formulation of the action-value function $Q\Big(\textbf{s}_{k},\textbf{a}_{k}\Big)$ holds the same as in \eqref{QRL}. However, the action-value function in \eqref{QRLcombined} incorporates the MPC objective costs in $r_{k+1}$ and the moving $M$ and prediction $N$ horizons in $\gamma$: 
\begin{align}\label{QRLcombined}
&\!\!\!\!\!\!\!\!\!Q\Big(\textbf{s}_{k},\textbf{a}_{k}\Big)\!\leftarrow\!Q\Big(\textbf{s}_{k},\textbf{a}_{k}\Big) \notag +\alpha\Big[ r_{k+1}\Big(\textbf{s}_{k+1},\textbf{a}_{k+1}\Big)\\& \qquad\;\; +\gamma \max_{\textbf{a}} Q\Big(\textbf{s}_{k+1},\textbf{a}\Big)\! -Q\Big(\textbf{s}_{k},\textbf{a}_k\Big) \Big].
\end{align}
From \eqref{QRLcombined}, the general form of bridging MPC and RL for fish growth is
 \begin{subeqnarray}\label{mpc_combine}
&&\!\!\!\!\!\!\!\!\!\!\!\!\!\!\!\!\!\!\!\!\!\!\displaystyle Q\Big(\textbf{s}_{k},\textbf{a}_{k}\Big)\leftarrow\ Q\Big(\textbf{s}_{k},\textbf{a}_{k}\Big) +\alpha\Big[-\!\! \int_{t_k}^{t_{k+1}}\!\!\!\!\!\ell(.) \der \tau \notag \\&& \qquad\qquad\qquad\,\,\,\,\, +\gamma \max_{\textbf{a}} Q\Big(\textbf{s}_{k+1},\textbf{a}\Big)\!
\notag \\ && \qquad\qquad\qquad\,\,\,\,-Q\Big(\textbf{s}_{k},\textbf{a}_k\Big) \Big] \slabel{eq-mpc11a-a}\\ 
&&\!\!\!\!\!\mbox{s.t}\quad \dot{\tilde{w}}(t_k)= g\big(\tilde{w}(t_k),u(t_k)\big) \slabel{eq-mpc1b-a}\\
&&\!\!\!\!\! u_{\mbox{\scriptsize min}}\leqslant u(t) \leqslant u_{\mbox{\scriptsize max}}, \quad \forall t \in[t_k,\, t_{k+1}]\slabel{eq-mpc1c-a}\\
&&\!\!\!\!\! \Delta u(t_k)= u(t_k)-u(t_{k-1})\slabel{eq-mpc1cbis-a}\\
 & &\!\!\!\!\! w_0 \leqslant \tilde{w}(t_k) \leqslant w_{\mbox{\scriptsize end}} , \quad \forall t \in[t_k,\, t_{k+1}]\slabel{eq-mpc1c0-a} \\
&&\!\!\!\!\! \tilde{w}(t_k)=w(t_k), \quad \tilde{w}(0) =w(t_0) \slabel{eq-mpc1d-a},
  \end{subeqnarray}
where $g$ represents the identified or calibrated model derived with the assumptions in the fish growth model \eqref{sys1}, as mentioned in Section~\ref{sub-bioenergetic}.
Table \ref{sample-table} lists the features of learning the MPC cost from RL and the RL reward from MPC. 

\section{Conclusion and future outlook}\label{conclusion}
Aquaculture systems are very challenging and complex due to the feeding and water quality control demands crucial to optimizing the growth and survival of fish. Thus, model-based and model-free feedback control systems ensure better feeding and water quality monitoring for better production. Model-based (MPC) and model-free RL are essential techniques demonstrating complementary lines in their approach, optimality, and constraint satisfaction. The synergy of MPC and RL methods is full of potential for the control theory and machine-learning communities. The complementarity lies in an intuitive description and conceptual reflection on their central concepts. This paper addresses the model-based and model-free control design techniques for fish growth in aquaculture systems, primarily algorithms that optimize the feeding and water quality of the dynamic fish growth process. Specifically, we addressed all aquaculture backgrounds required to understand the challenges, such as the effects of feeding and water quality on growth and survival, the challenges for building fish growth models, and the constraint satisfaction challenge related to RL. Solutions that combine MPC and RL display promising benefits for efficiently handling constraint satisfaction and determining better trajectories and policies from value-based RL.

The future of fully regulated and autonomous intelligent aquaculture systems has exciting benefits and potential, with many challenges before its realization. Therefore, aquaculturists and engineers have many relevant opportunities to work at the synergy level of MPC and RL and be part of one of the essential engineering developments and scientific achievements for living aquaculture fish in tanks and open-cage systems.

\section{Appendix}\label{App}
The effects of temperature $\tau(T)$, un-ionized ammonia $v(UIA)$, and dissolved oxygen $\sigma(DO)$ on food consumption are described, respectively \cite{Yan:98}:
\begin{equation*}
\tau(T)=
\left\{\begin{array}{llll}
\displaystyle \exp{\left\{-\kappa\Big(\dfrac{T-T_{opt}}{T_{\mbox{\tiny max}}-T_{opt}}\Big)^4\right\}} \quad \mbox{if}\quad T>T_{opt},\\
\exp{\left\{-\kappa \Big(\dfrac{T_{opt} -T}{T_{opt}-T_{\mbox{\tiny min}}}\Big)^4\right\}} \quad \mbox{if}\quad T<T_{opt},
\end{array}\right.
\end{equation*}
where $\kappa=4.6$, and
\begin{equation*}
v(UIA)\!\!=\!\!
\left\{\begin{array}{llll}
\displaystyle 1 \qquad\qquad\qquad~\, \mbox{if} \quad UIA<UIA_{\mbox{\tiny crit}},\\
\!\!\dfrac{UIA_{\mbox{\tiny max}}-UIA}{UIA_{\mbox{\tiny max}} -UIA_{\mbox{\tiny crit}}} \, \mbox{if}\, UIA_{\mbox{\tiny crit}}< UIA< UIA_{\mbox{\tiny max}},\\
0 \qquad\qquad\qquad\quad \mbox{elsewhere}.
\end{array}\right.
\end{equation*}
\begin{equation*}
\sigma(DO)=
\left\{\begin{array}{llll}
\displaystyle 1 \qquad\qquad\qquad \mbox{if} \quad DO>DO_{\mbox{\tiny crit}},\\
\!\!\dfrac{DO - DO_{\mbox{\tiny min}}}{DO_{\mbox{\tiny crit}} - DO_{\mbox{\tiny min}}}~\, \mbox{if}\quad DO_{\mbox{\tiny min}}\!<\!DO\!<\!DO_{\mbox{\tiny crit}},\\
0 \qquad\qquad\qquad \mbox{elsewhere}.
\end{array}\right.
\end{equation*}

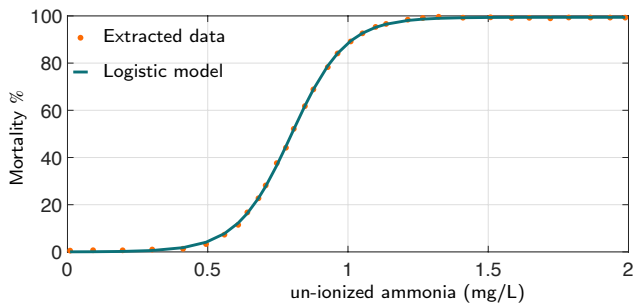
\begin{figure}[!h]
 \vspace{0.3cm}
\centering
   \begin{overpic}[scale =0.18]{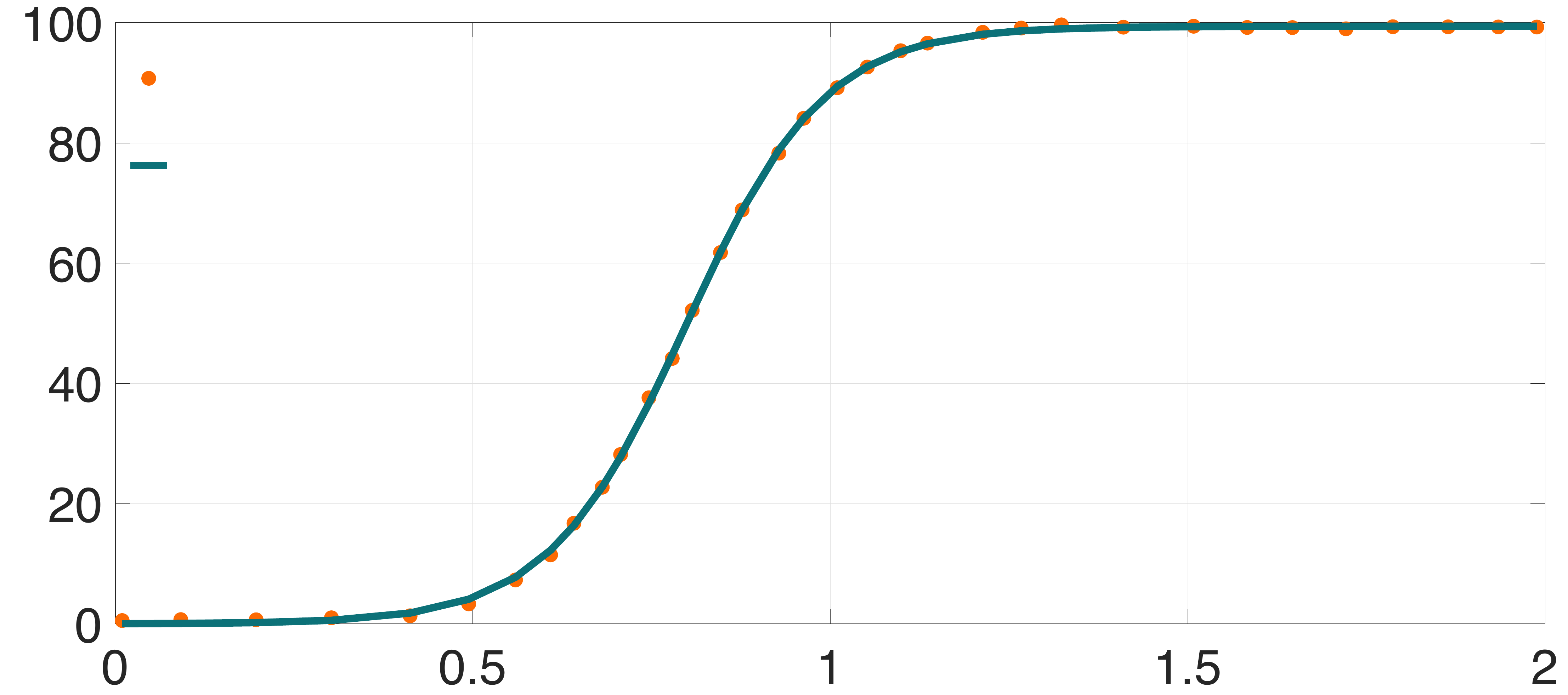}
   \end{overpic}
   \put(-235,35){\scriptsize \rotatebox{90}{Mortality $\%$}}
   \put(-200,89){\scriptsize Extracted data}
   \put(-200,76){\scriptsize Logistic model}
   \put(-130,-7){\scriptsize un-ionized ammonia (mg/L) }
   \caption{Relation between fish mortality coefficient and un-ionized ammonia}\label{UIA_mortality} 
 \end{figure}

The fish mortality coefficient $k_1$ depends on the un-ionized ammonia ($UIA$) factor, which has a logistic regression form, as follows: 
\begin{equation*}
  k_1(UIA) = \frac{\mathcal{Z}}{1 + \exp\{-\beta(UIA - \eta)\}},
\end{equation*}
where $\mathcal{Z}$, $\beta$, and $\eta$ are the three tuning parameters in the logistic fitting. The fish mortality points were extracted from a real experiment in \cite{Sink2010}, and the three tuning parameters were $\mathcal{Z}=99.41$, $\beta=10.36$, and $\eta=0.80$. Figure~\ref{UIA_mortality} illustrates the extracted points and logistic fitting.
\begin{table}[!t]
\begin{scriptsize}
\caption{Nomenclature and main parameters of the growth model}
\begin{center}
\begin{tabular}{| c || c || c |}
 \hline %
\textbf{~Symbol~} & \textbf{~~~~Description~~~~} & \textbf{~~~~Units~~~~} \\ \hline
$\xi$ & Total fish biomass & \si{kcal \per pond} \\
$p$ & Total number of fish & \si{fish \per pond} \\
$p_s$ & Stocking fish number & \si{fish \per day\per pond}\\
$\xi_i$ & Individual fish biomass during fish stocking & \si{kcal \per fish} \\
$\xi_a$ & Mean fish biomass & \si{kcal \per pond}\\
$k_1$ & Fish mortality profile & - \\
~~~$m$~~~ &Exponent of body weight for net anabolism &{$0.67$}\\
~~$t$~~ & Time & \si{day} \\ 
$n$ & Exponent of body weight for fasting catabolism &~~~~{$0.81$}~~~~ \\  \hline
$f$ & Relative feeding rate &$0\!<\!f\!<\!1$ \\ 
$T$ &~~Temperature &\si{^0}\si{C} \\
$DO$ & Dissolved oxygen & \si{mg/l} \\ 
$UIA$ &~Un-ionized ammonia~ & \si{mg/l} \\  \hline
$b$ & Efficiency of food assimilation & {$0.62$} \\ 
$a$ & Fraction of the food assimilated &{$0.53$}\\ 
$h$ &Coefficient of food consumption &$0.8$ \si{kcal^{1\!-\!m}}\si{/day}\\ 
$k_{\mbox{\tiny min}}$ &~Coefficient of fasting catabolism~ &$0.00133$\si{N}\\ 
$j$ &~Coefficient of fasting catabolism~ &0.0132\si{N}\\ 
$T_{opt}$ & Optimal average water temperature level &$33^0$\si{C}\\
$T_{\mbox{\tiny min}}$& Minimum temperature level &$24^0$\si{C}\\
$T_{\mbox{\tiny max}}$ & Maximum temperature level &$40^0$\si{C}\\ 
$UIA_{\mbox{\tiny crit}}$ &~~~Critical un-ionized ammonia limit~~~ &$0.06$\si{mg/l}\\
$UIA_{\mbox{\tiny max}}$ & Maximum un-ionized ammonia level &$1.4$\si{mg/l}\\
$DO_{\mbox{\tiny crit}}$ & Critical dissolved oxygen limit&$0.3$\si{mg/l}\\ 
$DO_{\mbox{\tiny min}}$ & Minimum dissolved oxygen level &$1$\si{mg/l}\\ 
$r$ & Daily ration &\si{kcal/day}\\
$R$ & Maximal daily ration & $10\%$ \si{BWD}\\
$BWD$ & Average body weight per day &\si{kcal/day}\\ \hline 
$\tau$ &~Temperature factor~ &$0\!<\!\tau\!<\!1$ \\ 
$\sigma$ & Dissolved oxygen factor & $0\!<\!\sigma\!<\!1$ \\ 
$v$ & Un-ionized ammonia factor & $0\!<\!v\!<\!1$ \\ 
$\rho$ & Photoperiod factor & $0\!<\!\rho\!<\!2$\\ \hline 
\end{tabular}
\end{center}
\label{para}
\end{scriptsize} 
\end{table}

\bibliography{biblio}

\end{document}

%% file: setup-IEEEtrans.tex
\renewcommand{\intextsep}{0pt} 

\allowdisplaybreaks 

\def\aujour{\number\day \space \ifcase\month\or
janvier\or f�vrier\or mars\or avril\or mai\or
juin\or juillet\or ao�t\or septembre\or octobre\or
novembre\or d�cembre\fi \space \number\year}

\def\cH{{\cal H}}

\def\cL{{\cal L}}

\newtheorem{remark}{Remark}

\def\C{{\setbox0=\hbox{$\displaystyle{\rm C}$}
        \hbox{\hbox to0pt{\kern 0.4\wd0\vrule height 0.95\ht0\hss}\box0}}}
\def\Q{{\setbox0=\hbox{$\displaystyle{\rm Q}$}%
    \hbox{\raise 0.2\ht0\hbox to0pt{\kern 0.4\wd0\vrule height
    0.85\ht0\hss}\box0}}} 
\def\R{\mathop{\rm I\mkern -3.5mu R}} 

\def\cH2{{\cal H}_2} 

\def\cL2{\mathop{\mathcal L}_{2}} 

\def\cRH2{\mathop{\cal R \cal H}_2} 
\def\cRL2{\mathop{\cal R \cal L}_{2}} 

\DeclareMathOperator*{\der}{d}

\makeatletter
\DeclareRobustCommand\sfrac[1]{\@ifnextchar/{\@sfrac{#1}}
                                            {\@sfrac{#1}/}}
\def\@sfrac#1/#2{\leavevmode\kern.1em\raise.5ex
         \hbox{$\m@th\fontsize\sf@size\z@
                           \selectfont#1$}\kern-.1em
         /\kern-.15em\lower.25ex
          \hbox{$\m@th\fontsize\sf@size\z@
                            \selectfont#2$}}
\makeatother
